\documentclass[a4paper,11pt,twocolumn,accepted=2020-10-24]{quantumarticle} 
\pdfoutput=1
\usepackage{amsmath,amsthm,amsfonts,amssymb,braket}
\usepackage{graphicx}
\usepackage{color}
\usepackage{subfigure}
\usepackage{enumitem}
\usepackage{booktabs}

\usepackage[numbers,sort&compress]{natbib}
\usepackage[colorlinks=true,citecolor=blue,linkcolor=blue]{hyperref}
\newcommand{\secref}[1]{\hyperref[#1]{{Section~\ref*{#1}}}}
\newcommand{\figref}[1]{\hyperref[#1]{{Fig.~\ref*{#1}}}}
\newcommand{\tabref}[1]{\hyperref[#1]{{Tab.~\ref*{#1}}}}
\newcommand{\appref}[1]{\hyperref[#1]{{Appendix~\ref*{#1}}}}
\newcommand{\eqnref}[1]{\hyperref[#1]{{(\ref*{#1})}}}
\newcommand{\claimref}[1]{\hyperref[#1]{{Claim~\ref*{#1}}}}

\newtheorem{theorem}{Theorem}
\newtheorem{lemma}[theorem]{Lemma}
\newtheorem{claim}[theorem]{Claim}

\begin{document}

\title{Optimization of the surface code design for Majorana-based qubits}

\author{Rui Chao}
\affiliation{University of Southern California, Los Angeles, CA, USA}
\author{Michael E. Beverland}
\author{Nicolas Delfosse}
\author{Jeongwan Haah}
\affiliation{Microsoft Quantum and Microsoft Research, Redmond, WA, USA}

\begin{abstract}
The surface code is a prominent topological error-correcting code exhibiting high fault-tolerance accuracy thresholds. 
Conventional schemes for error correction with the surface code place qubits on a planar grid and assume native CNOT gates between the data qubits with nearest-neighbor ancilla qubits.

Here,
we present surface code error-correction schemes using {\em only} Pauli measurements on single qubits and on pairs of nearest-neighbor qubits. 
In particular, we provide several qubit layouts that offer favorable trade-offs between qubit overhead, circuit depth and connectivity degree.
We also develop minimized measurement sequences for syndrome extraction, enabling reduced logical error rates and improved fault-tolerance thresholds. 

Our work applies to topologically protected qubits realized with Majorana zero modes and to similar systems in which multi-qubit Pauli measurements rather than CNOT gates are the native operations.
\end{abstract}

\maketitle

\section{Introduction}

Fault tolerance is widely believed to be necessary to run viable applications on a quantum computer.
Errors occurring during the computation must be corrected at regular intervals and faster than they accumulate.
The design of a fault-tolerant quantum computer is constrained by the limitations of quantum hardware.
For instance, at present it remains extremely challenging to produce a large number of high-quality qubits. 
Moreover, quantum chips generally offer only a reduced qubit connectivity, often limited to nearest-neighbor interactions. 

Given these constraints, the surface code~\mbox{\cite{kitaev2003surfacecode, dennis2002surfacecode, fowler2012surfacecode}} has proven to be one of the leading candidates for error correction in a quantum computer.
Two crucial properties make the surface code very attractive for a first generation of fault-tolerant quantum computers:
(i) Error correction with the surface code can be implemented on a planar grid of qubits using only single-qubit operations and nearest-neighbor gates,
(ii) The surface code tolerates qubits and elementary operations affected by relatively high error rates~\cite{raussendorf2007surfacecode, fowler2009high_threshold}. 
These properties have been established for qubits equipped with CNOT gates, {\em e.g.},  superconducting qubits; however, it is unclear whether similar results hold with other types of qubits.

In this article, we consider the performance of the surface code for measurement-based qubits. 
These qubits do not possess a native CNOT: instead they are equipped with single-qubit and two-qubit Pauli measurements. 
These two sets of operations, based on CNOT gates or Pauli measurements, are equivalent in the sense that they can simulate each other in polynomial time. 
In particular, the surface code can be implemented with measurement-based qubits up to a polynomial overhead.
However, for practical purposes a polynomial overhead can have dramatic consequences.
A naive translation from the CNOT-based implementation of the surface code error correction into a measurement-based circuit leads to a blow-up of the qubit overhead. 
Five times as many ancilla qubits are required, since each CNOT gate implemented as a sequence of measurements requires an extra ancilla qubit.
More qubits also incur more potential fault locations, which result in a significant reduction of the surface code performance, and which may cancel property (ii).

In this work, we propose implementations of surface code error correction with measurement-based qubits that retain both of the positive properties (i) and (ii) described above, and meanwhile (iii) consume the same number of ancilla qubits as the CNOT-based implementation.
Note that property (iii) is valuable, since the extra ancillas required for emulating CNOT gates is one of the main potential drawbacks of measurement-based qubits. 
Our implementations rely on two main ingredients.
First, we design planar qubit layouts for the surface code, where the ancilla qubits can be recycled both for emulating the CNOT gates and for revealing the syndrome bits, via only local measurements. 
Second, we optimize the decomposition of the CNOT-based circuit into measurements by reducing the circuit depth, enabling a shorter error-correction cycle. 
By reducing the number of locations at which faults can occur in each error-correction cycle, this also leads to a reduction of the logical error rate. 

We numerically simulate the error-correction schemes which combine the optimized layouts and syndrome-extraction circuits, using the Union-Find decoder~\cite{delfosse2017peeling, delfosse2017unionfind}. 
Under a circuit-level error model where each location experiences depolarizing noise, we observe empirical error rate thresholds as high as $2.37\times 10^{-3}$. 

One potential application of our measurement-based surface code designs is for quantum computers consisting of Majorana zero modes~\cite{karzig2017majorana}, where physical qubits are encoded into an even number of Majoranas with fixed parity. 
Both the storage and manipulation with Majorana-based qubits are topologically protected, {\em i.e.}, robust to local perturbations.  
In particular, reliable measurements of qubit Pauli operators---which can be realized by gathering relevant constituent Majoranas and measuring their joint parities---are  amenable to our schemes.  
Thanks to the topological protection, one should be able to manufacture high-quality Majorana qubits with error rates far below the thresholds required to implement our schemes.
Let us remark that there have been studies on the implementations of the Bacon-Shor code~\cite{Knapp2018modeling}, Majorana surface codes~\cite{li2016noise,plugge2016roadmap} and Majorana color codes~\cite{litinski2017combining}.
These works mostly assume that plaquette operators with weight four or six are \emph{directly} measured without the usual need of ancilla qubits. 
Here, we instead allow the use of ancillas and restrict to at most weight-two Pauli measurements, mainly to avoid harmful correlated noise or reduction of the effective distance when considering circuit-level errors. 
It is also likely that achieving high-fidelity measurements of more than two qubits will prove much more difficult in practice.

\secref{sec:measbased} introduces notions about measurement-based qubits along with a simplified noise model. 
\secref{sec:windmill} presents a windmill-like qubit layout which has the same qubit overhead as in  the standard CNOT-based surface code error correction. 
In addition, \secref{sec:alterlayout} gives two alternative layouts that feature favorable circuit depth and qubit connectivity, respectively. 
\secref{sec:cnotnot} explains how to measure the weight-four plaquette operators using fewer time steps than the naive translation from CNOT gates. 
\secref{sec:numerics} lists the results of the numerical simulations and gives threshold estimates. 
We introduce a mapping of error distributions which expedites the sampling of errors in simulation significantly.
The resulting error distribution, called the inclusive error model as explained in \appref{app:inclusivemodel},
is equivalent to the conventional error model ({\em e.g.,} the depolarizing noise or the bit flip noise) in all important regimes
and may be of independent interest.

\section{Measurement-based qubits}
\label{sec:measbased}

We consider a set of qubits equipped with {\em single-qubit measurements} of Pauli matrices $X, Y$ and $Z$. 
The only available entangling operations are {\em joint measurements}---measurements of two-qubit Pauli operators acting on connected qubits. 
A $T$ gate or another non-Clifford gate must be added to the gate set in order to achieve universality. 
The optimization of the production of non-Clifford operations is not considered in this work. 
We focus on the design of the error-correction schemes  that depend only on the Clifford part of the gate set---single-qubit measurements and joint measurements.

A graph, whose edges support joint measurements, describes the qubit connectivity. 
In order to make the chip design possible, it is often necessary to restrict ourselves to low connectivity (low degree) and to graphs that can be embedded in a plane with a small number of crossing edges. 
Planar connectivity graphs are optimal in that regard.

In addition to single-qubit and joint measurements, measurement-based qubits are equipped with an additional operation that we call a Pauli update and that can be implemented in the classical control device without any physical action on the qubits.
In general, each measurement $M_P$ is followed by a Pauli update $U_Q$ which applies the Pauli operation $Q$ to the system if and only if the outcome of the measurement of Pauli operator $P$ is non-trivial.

\begin{figure}
\centering
\hspace{-.5cm}\includegraphics[scale=.6]{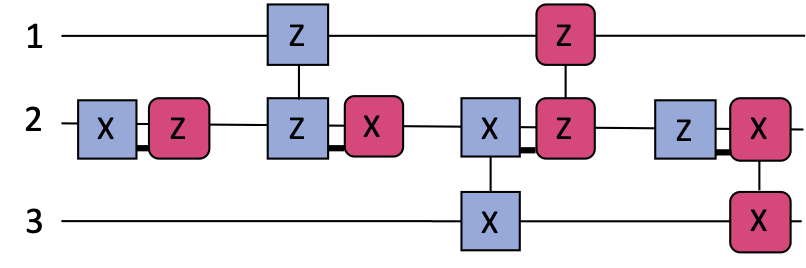}
\caption{
A CNOT gate with control qubit 1 and target qubit 3 using qubit 2 as an ancilla. 
The CNOT gate is implemented as a sequence of two single-qubit measurements and two joint measurements. 
Each measurement $M_P$ (represented by a Pauli in blue squares, with joint measurements connected with vertical lines) is followed by a Pauli update $U_Q$ (represented by a Pauli in rounded pink squares, connected to the measurement with a thick horizontal line). 
The update $U_Q$ applies the Pauli operation $Q$ if the outcome of the preceding measurement is non-trivial.
}\label{fig:cnot}
\end{figure}

Sequences of Pauli measurements and Pauli updates can generate arbitrary Clifford circuits.
The behavior of these circuits can be verified by keeping track of the stabilizers of the state after each measurement and update. 
A CNOT gate implementation for measurement-based qubits is given in \figref{fig:cnot}.
An ancilla qubit is necessary in order to implement this two-qubit gate.

\begin{figure}[!b]
\centering
\includegraphics[scale=.6]{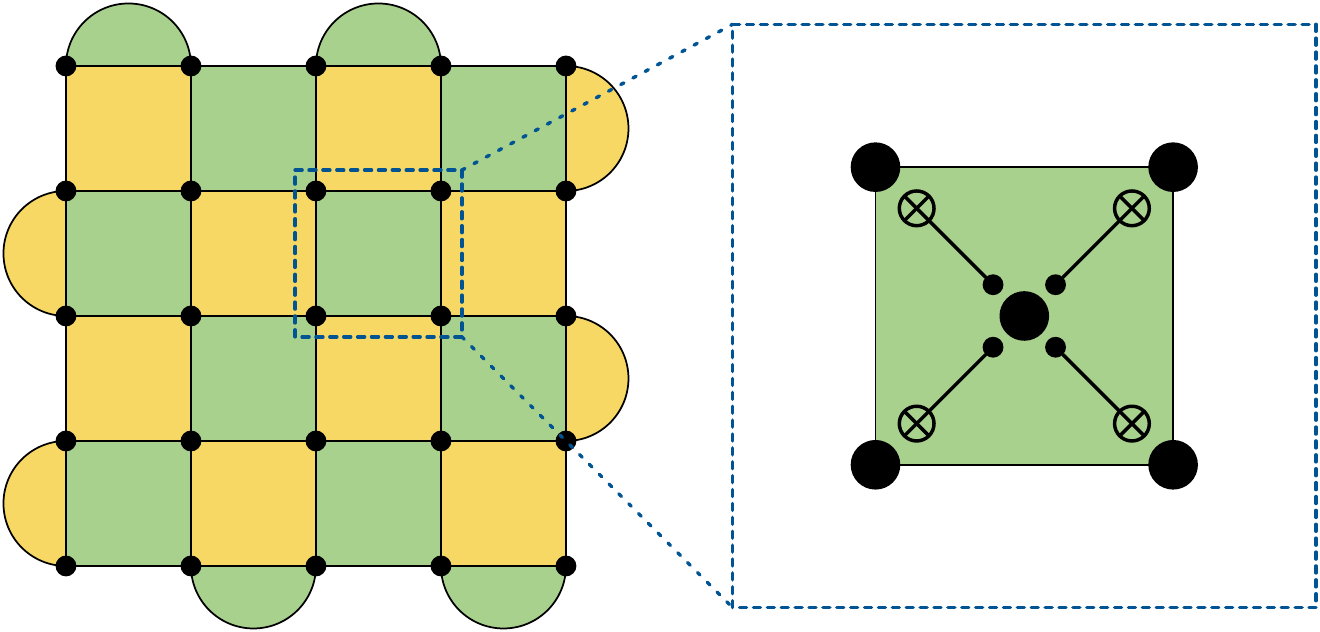}
\caption{
A distance-five surface code encoding 1 logical qubit into 25 physical qubits. 
Black nodes represent physical qubits and each colored region corresponds to the measurement of a syndrome bit. 
Green and yellow plaquettes support respectively $X$-type and $Z$-type measurements. 
With CNOT-based qubits, a measurement is implemented locally inside a plaquette using one ancilla qubit connected to the plaquette qubits by CNOT gates. 
}\label{fig:surface_code}
\end{figure}

\begin{figure*}
\centering
\begin{tabular}{ccc}
\subfigure[\label{}]{
\raisebox{.5cm}{\includegraphics[scale=.75]{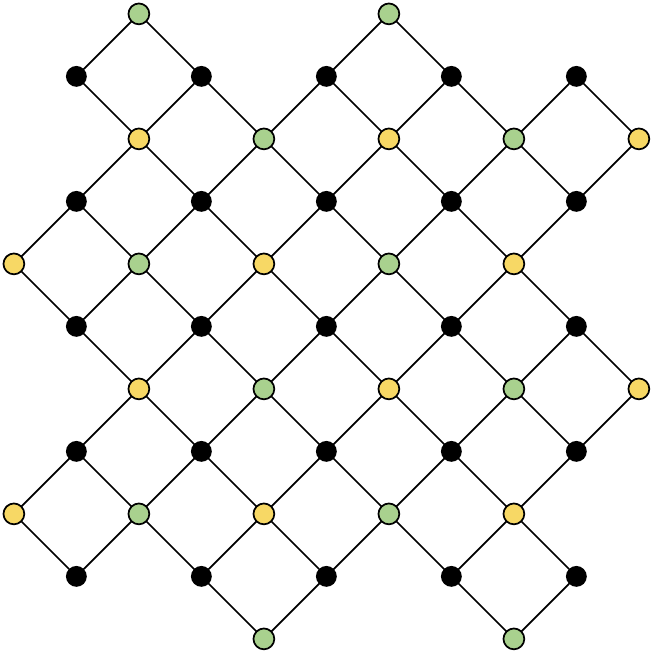}}}
&\quad
\subfigure[\label{fig:naive}]{
\raisebox{.5cm}{\includegraphics[scale=.75]{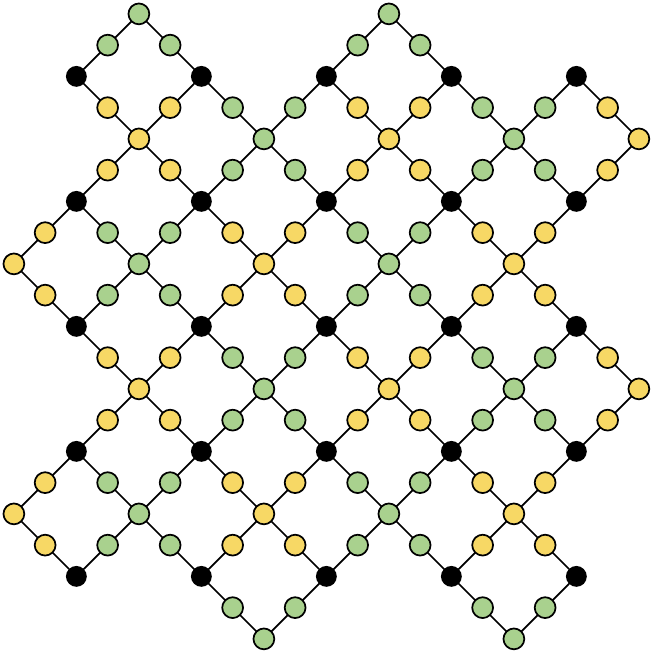}}}
&\quad
\subfigure[\label{}]{
\raisebox{.5cm}{\includegraphics[scale=.75]{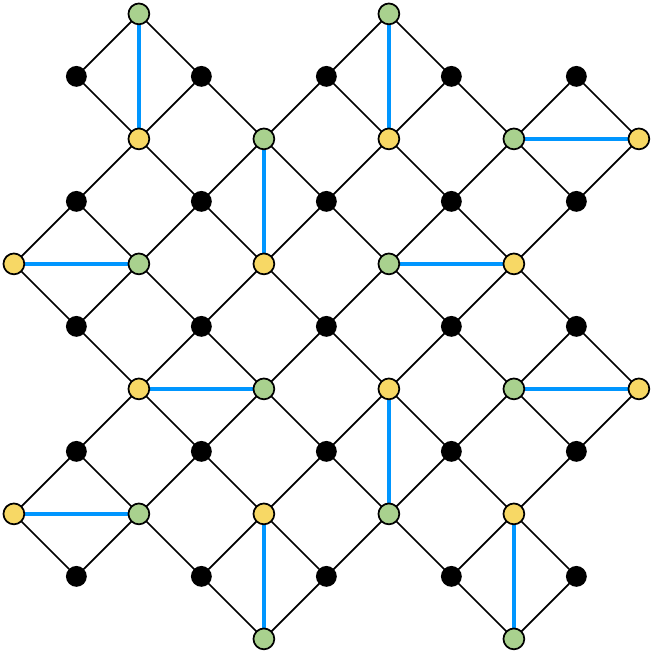}}}
\end{tabular}
\caption{
Connectivities required for three different implementations of the distance-five surface code.
Black nodes represent data qubits and colored nodes are ancilla qubits consumed by the syndrome-extraction circuits. 
All the multi-qubit operations, {\em i.e.}, CNOT gates or joint measurements, are supported on the links of the graphs.
(a) Connectivity required for a CNOT-based implementation of the surface code.
(b) The naive conversion of each CNOT into a product of single-qubit measurements and joint measurements costs an extra ancilla for each CNOT link.
(c) Optimized layout implementing the surface code with measurement-based qubits.
Only one ancilla qubit is consumed per plaquette at the price of one extra link per ancilla.
The windmill layout (c) is described further in \figref{fig:windmill}.
}\label{fig:layout}
\end{figure*}

\subsection{Noise model}
\label{sec:noisemodel}

We assume a circuit-level noise model where each elementary operation in the error-correction  circuit is afflicted by a fault, chosen according to some distribution, from a certain finite set specified as follows. 
\begin{itemize}
\item Single-qubit identity gate faults:
\[ \{I, X, Y, Z\}. \]
\item Single-qubit measurement faults:
\[ \{I, X, Y, Z\} \times \{\textrm{flip, no flip}\}. \]
\item Joint measurement faults:
\[ \{I, X, Y, Z\} ^{ \otimes 2} \times \{\textrm{flip, no flip}\}. \]
\end{itemize}
Here, ``flip'' or ``no flip'' indicates whether or not the measurement outcome is incorrectly flipped, that is, whether or not an erroneous Pauli update is introduced.  
Clearly, this is essentially equivalent to a stochastic Pauli error model. 

In all the numerical simulations that we have performed (to be explained in \secref{sec:numerics}), individual elementary operations are faulty independently with same probability.  
When faulty, an operation is affected by a fault which is chosen from the set of all possible nontrivial faults, uniformly at random. 

Although our noise model resembles the depolarizing error model typically assumed for  the CNOT-based circuits, it is difficult to establish a meaningful or quantitative comparison between these two models and thus the corresponding error-correction schemes. 

\section{Surface code layout}
\label{sec:windmill}

The surface code~\cite{kitaev2003surfacecode, dennis2002surfacecode, fowler2012surfacecode} encodes one logical qubit into a grid of $d \times d$ physical qubits as shown in \figref{fig:surface_code}. 
Additional qubits are consumed by the implementation of the correction scheme. 
Error correction is based on the measurements of the plaquette operators of the form $X_{q_1} X_{q_2} X_{q_3} X_{q_4}$ acting on the four qubits of green plaquettes or $Z_{q_1} Z_{q_2} Z_{q_3} Z_{q_4}$ over yellow plaquettes.
Side plaquettes involve only two qubits.
A round of stabilizer measurements produces an outcome bit for each plaquette, the so-called {\em syndrome} bits.
The decoder provides an estimation of the errors which occur based on the knowledge of the syndrome. 
A number of efficient decoding algorithms have been proposed for the surface code~\cite{dennis2002surfacecode}. 
In this work, we consider the Union-Find decoder for its rapidity~\cite{delfosse2017peeling, delfosse2017unionfind}.

\medskip

\begin{figure*}
\centering
\includegraphics[scale=.5]{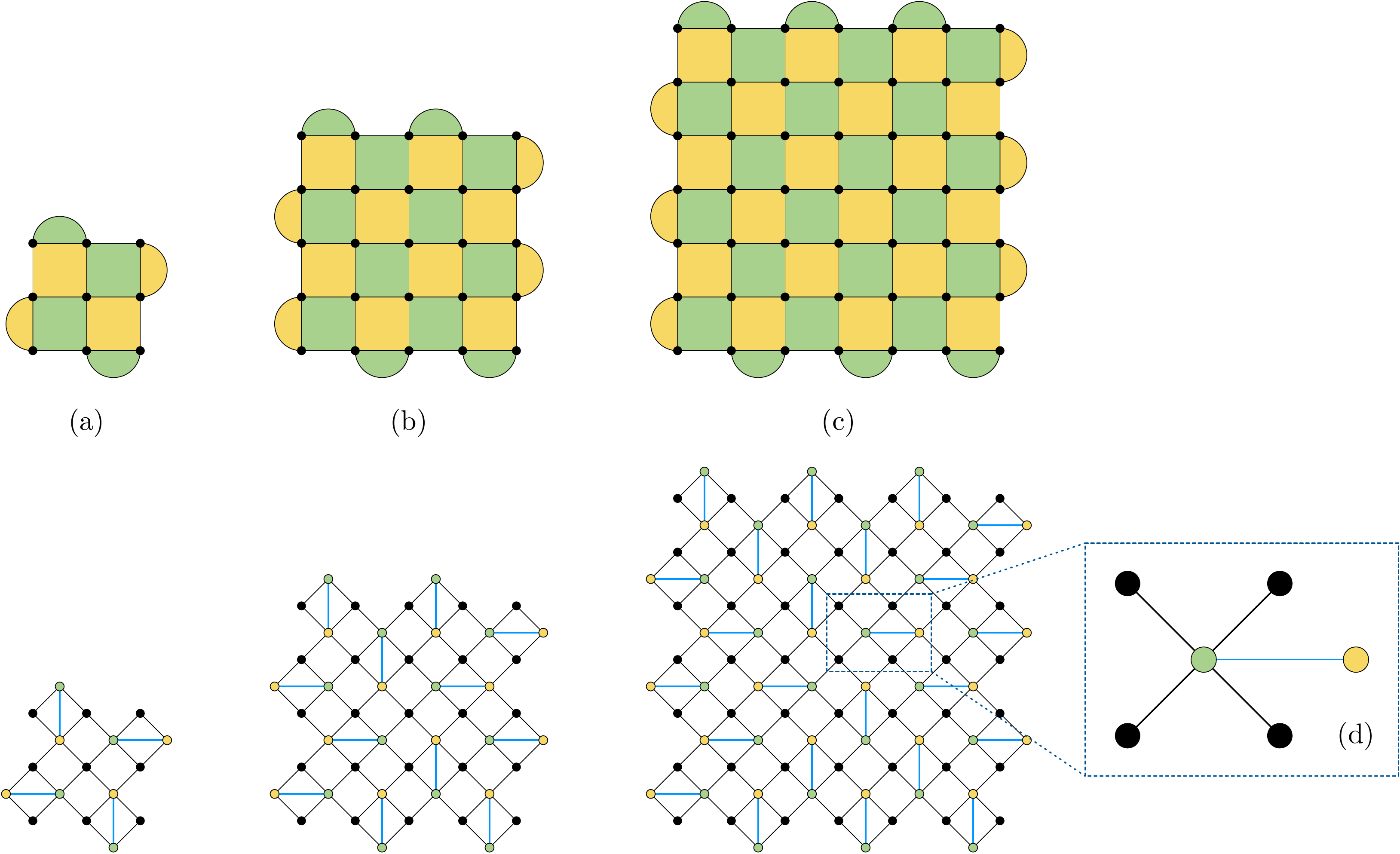}
\caption{
The windmill layout implementing the surface code syndrome-extraction circuit via measurement-based qubits for distance 3~(a), 5~(b) and 7~(c). 
Blue edges indicate the additional connectivity in comparison with the CNOT-based layout. 
A single ancilla per plaquette is sufficient.  
Ancilla qubits have a degree-five connectivity and data qubits remain degree-four.
In order to implement a plaquette measurement, we need a pair of ancillas.
We use the ancilla in the center of the plaquette and its neighbor ancilla (linked by a blue edge).
A complete round of measurements is done in two steps, measuring together all green plaquettes and then all yellow plaquettes.
In~(d) we show the connectivity used during the measurement of a greeen plaquette.
The neighbor yellow ancilla is required.
}\label{fig:windmill}
\end{figure*}

\begin{figure*}
\centering
\begin{tabular}{cc}
\subfigure[\label{fig:double_ancilla_layout}]{
\raisebox{.5cm}{\includegraphics[scale=.9]{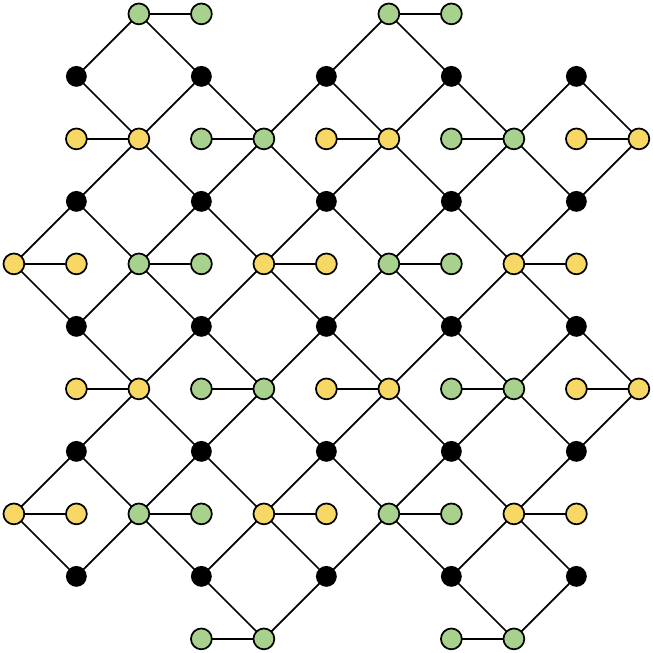}}}
&\qquad \qquad
\subfigure[\label{fig:twoandhalf_ancilla_layout}]{
\raisebox{.5cm}{\includegraphics[scale=.9]{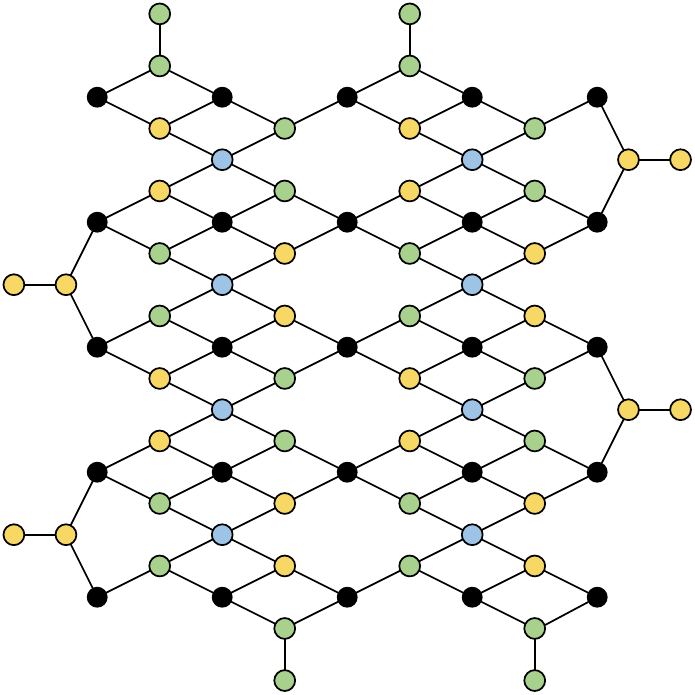}}}
\end{tabular}
\caption{
(a) Double ancilla layout for the distance-five surface code. 
Two ancilla qubits are used for each plaquette. 
The measurement depth can be reduced by a factor of two in comparison with the windmill layout, at the price of consuming twice as many ancilla qubits.
(b) 
A layout for the distance-five surface code with 
a connectivity graph of maximum degree~4, {\em i.e.}, each qubit allows for joint measurements 
with at most 4 different neighboring qubits.
Green ancillas are for $X$-type stabilizers and yellow ancillas are for $Z$-type stabilizers.
Blue ancillas are shared between two types of plaquettes.
There are approximately~$2.5$ ancillas per plaquette.
}
\end{figure*}

\begin{figure*}
\centering
\begin{tabular}{c}
\subfigure[\label{fig:naiveX4}]{
\raisebox{0cm}{\includegraphics[scale=.55]{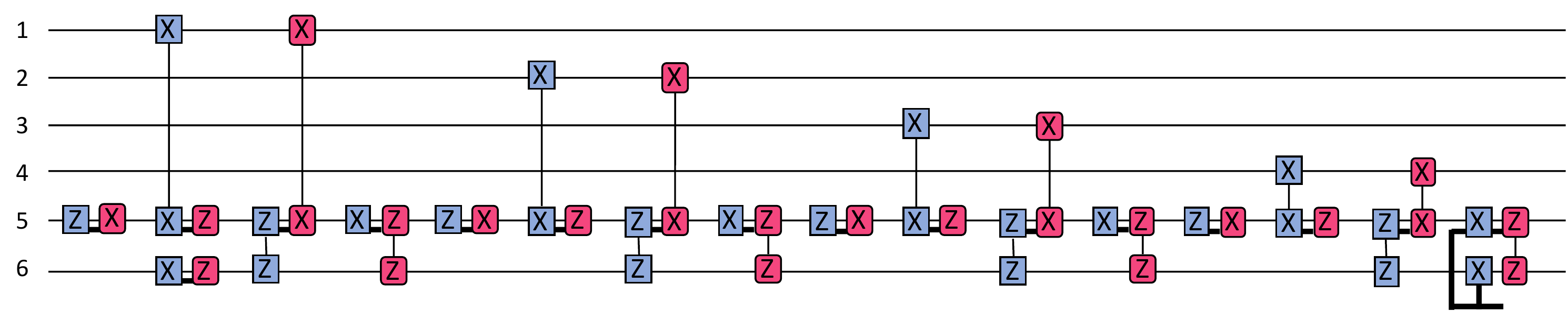}}}
\\
\subfigure[\label{fig:compressedX4}]{
\raisebox{.2cm}{\includegraphics[scale=.55]{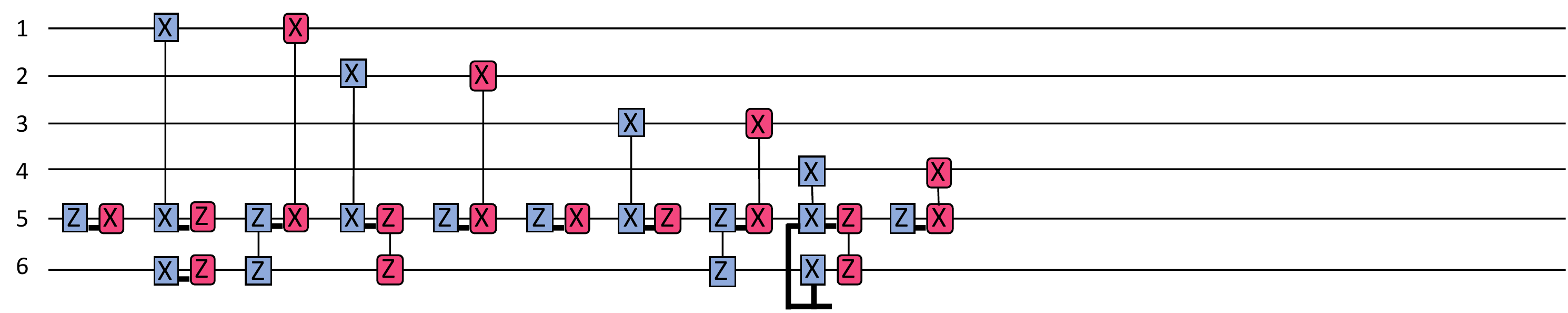}}}
\end{tabular}
\caption{
Measurement-based circuits which implement the $X^{\otimes 4}$ stabilizer measurement. 
(a) This gadget is built from four uses of the CNOT gadget in \figref{fig:cnot}. 
It takes 16 time steps, involving 10 single-qubit and 8 joint measurements.
(b) This (significantly more efficient) gadget has been found by exhaustive search of measurement-based sequences. 
It takes 10 time steps, involving 5 single-qubit and 6 joint measurements. 
For both (a) and (b), the measurement outcome of $X ^{\otimes 4}$ is obtained from the parity of those of a pair of measurements, indicated by a dangling  junction of thick lines.
}\label{fig:X4}
\end{figure*}

Qubits equipped with native CNOT gates consume exactly one ancilla qubit per syndrome bit extracted during a round of syndrome measurement. 
Figure~\ref{fig:surface_code} shows the locations of ancilla qubits and the connectivity of the CNOT gates used inside a plaquette.
Overall, a square grid connectivity is enough to implement the surface code with CNOT-based qubits.
Figure~\ref{fig:layout} compares the CNOT connectivity graph with the connectivity required for measurement-based qubits. 
A naive solution is to simulate each CNOT by a sequence of measurements. 
This costs one extra ancilla qubit per CNOT as we can see in \figref{fig:naive}. 
The number of ancillas jumps by a factor of five for large minimum distance. 
The implementation of the smallest surface code (with distance three) would require 33 qubits instead of 17 qubits.

\medskip

The extra ancilla required for simulating a CNOT with measurement-based qubits cannot be omitted but it can be shared between multiple CNOT gates in the same plaquette and  between neighbor plaquettes. 
This leads to the \emph{windmill layout} described in \figref{fig:windmill}.
A plaquette is measured using two ancillas, one that stores the measurement outcome and a second one that supports the CNOT gates between the first ancilla and the four plaquette qubits.
This layout is particularly advantageous when the first priority is to minimize the qubit overhead, that is the number of physical qubits per logical qubit. 
Such priority 
brings two slight differences from the CNOT syndrome-extraction circuit.
First, the chip must allow for degree-five ancilla connectivity. 
This remains reasonable, perhaps at the price of a small increase of the measurement error rate.
The fact that the data qubits remain degree-four is encouraging.
Second, since the two connected ancillas are used together for a single plaquette measurement (see \hyperref[fig:windmill]{{Fig.~\ref*{fig:windmill}(d)}}), green and yellow plaquettes cannot be measured simultaneously. 
One has to implement a complete syndrome measurement round in two consecutive stages, each for one type of stabilizers. 

\subsection{Alternative layouts for syndrome extraction}
\label{sec:alterlayout}

We describe alternative layouts for syndrome extraction that may be useful in different regimes.

The windmill layout is designed to minimize the number of ancilla qubits. 
In the regime where it is easy to fabricate a large number of qubits, one may consider the {\em double ancilla layout} represented in \figref{fig:double_ancilla_layout}, which uses two ancilla qubits per plaquette. 
This reduces the time required for a complete error-correction cycle by a factor of two, but it costs twice as many ancillas as the windwill layout.

In an alternative layout depicted in \figref{fig:twoandhalf_ancilla_layout},  
all the physical qubits are connected with at most 4 neighboring qubits. 
This layout would be useful in a situation where a qubit may not be connected to 5 or more other qubits.
There are approximately $2.5$ ancillas per plaquette of the surface code.

\section{Optimization of the syndrome-extraction circuit}
\label{sec:cnotnot}

\begin{figure*}
\centering
\begin{tabular}{c@{\quad\quad}c}
\hspace{-.5cm}\subfigure[\label{fig:w_schedule}]{
\raisebox{.2cm}{\includegraphics[scale=1]{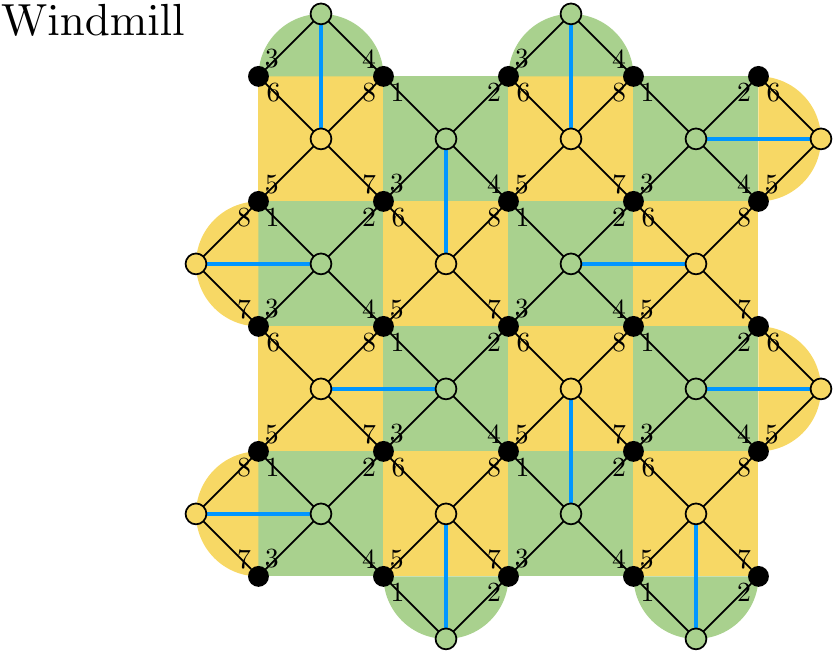}}}
&
\subfigure[\label{fig:d_schedule}]{
\raisebox{.2cm}{\includegraphics[scale=1]{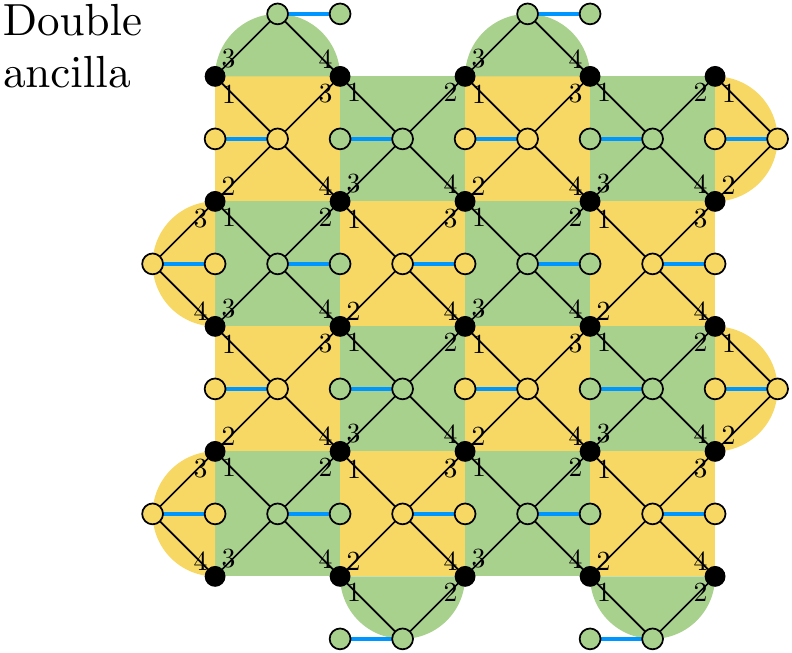}}}
\end{tabular}
\caption{Schedules for the joint Pauli measurements between data qubits and ancillas, using the (a) windmill layout in \figref{fig:windmill} and the (b) double ancilla layout in \figref{fig:double_ancilla_layout}.  
Solid lines indicate qubit connectivity. 
Bulk plaquettes are measured using gadgets as  in \figref{fig:compressedX4}, while boundary plaquettes are measured using gadgets as in \figref{fig:compressedX2}. 
Qubits of the $X$-type (green) and $Z$-type (yellow) stabilizers are acted upon in \texttt{Z} and \texttt{N} orders, respectively. 
The two types of stabilizers are measured in separate stages in~(a), and simultaneously in (b). 
}\label{fig:schedule}
\end{figure*}

Here we consider how to implement the weight-four $X^{\otimes 4}$ and $Z^{\otimes 4}$ stabilizer measurements required for the surface code using measurement-based qubits. 
The weight-two stabilizer measurements at the boundary of the lattice are implemented similarly.
This is a special case of a more general scheme for measuring arbitrary Pauli operators that we present in \appref{sec:optimizing-general-measurements}.

First let us review the approach already known for CNOT-based (rather than measurement-based) qubits~\cite{fowler2012surfacecode}. 
In that case, it is standard to include a single extra ancilla qubit which is entangled using CNOT gates with the four plaquette qubits which are to be measured jointly.
To implement an $X^{\otimes 4}$ joint measurement in CNOT-based qubits, one prepares the ancilla in $\ket{+}$, and sequentially applies a CNOT controlled by the ancilla and targeted on each of the four qubits involved in the measurement, before measuring the ancilla in the $X$ basis.
This circuit propagates error in a non-fault-tolerant manner.
For example, an $XX$ fault on the third CNOT gate will propagate to the final CNOT and will result in an $X$ error on each of the last two qubits in the measurement. 
However, it can be incorporated into a fault-tolerant error-correction protocol provided the errors resulting from single faults are sufficiently benign given the structure of the error-correction scheme.
In the case of the surface code, one can choose the ordering of the CNOT gates so that the weight-two error just described (an example of a {\em hook error}~\cite{dennis2002surfacecode}) 
is orthogonal to the minimum-weight logical operators, 
thereby behaving effectively as a weight-one error for the purposes of error correction~\cite{Fowler2010}.

Now let us turn to the case of measurement-based qubits. 
The simplest approach is to use precisely the same technique as is employed with CNOT-based qubits, 
but to decompose each CNOT in the circuit into measurements using the gadget shown in \figref{fig:cnot}.
This results in the circuit shown in \figref{fig:naiveX4}.
We remark that circuit \figref{fig:naiveX4} can be compressed by merging consecutive single-qubit $X$ or $Z$ measurements and accordingly changing the subsequent measurement bases.
The compressed circuit behaves the same as that in \figref{fig:naiveX4} in the absence of faults; however, it has malignant hook errors which the uncompressed circuit does not have. 

By performing a search of sequences involving single-qubit and joint  measurements, 
we have found small circuits that implement two- and three-target controlled-NOT gates 
$\mathrm{C}X\!X$ and $\mathrm{C}X\!X\!X$ (see \appref{sec:optimizing-general-measurements}). 
Iteratively using these circuits as modular components, 
one can measure a Pauli operator of arbitrary weight $n\ge2$ using two ancillas 
and either $n+2$ single-qubit measurements and $3n/2$ joint measurements when $n$ is even, 
or $n+4$ single-qubit measurements and $(3n+1)/2$ joint measurements when $n$ is odd.
Relevant to the surface code error correction is the special case of $n=4$,
as depicted in \figref{fig:compressedX4} for measuring bulk plaquettes. 
This sequence is significantly shorter and involves fewer measurements,
and thereby is expected to perform better than the naive circuit \figref{fig:naiveX4} built from CNOT gadgets.

Note that in gadget \figref{fig:compressedX4},
the two consecutive $Z$ measurements on qubit~5 in the middle may seem redundant, 
but are necessary to keep a single measurement error from propagating;
the latter $Z$ measurement on qubit~5 and the subsequent $X$ correction 
make sure that qubit~5 is set to the state~$\ket 0$,
even if the former $Z$ measurement was faulty.
Furthermore, even though it is a native operation with our measurement-based qubits 
to measure weight-2 plaquettes on the boundaries,
they should be measured using gadgets in \figref{fig:compressedX2},
rather than direct joint measurements, to make hook errors benign.

\begin{figure*}
\centering
\begin{tabular}{c@{$\quad\quad$}c}
\hspace{-.4cm}\subfigure[\label{fig:w_threshold}]{
\raisebox{.2cm}{\includegraphics[scale=.27]{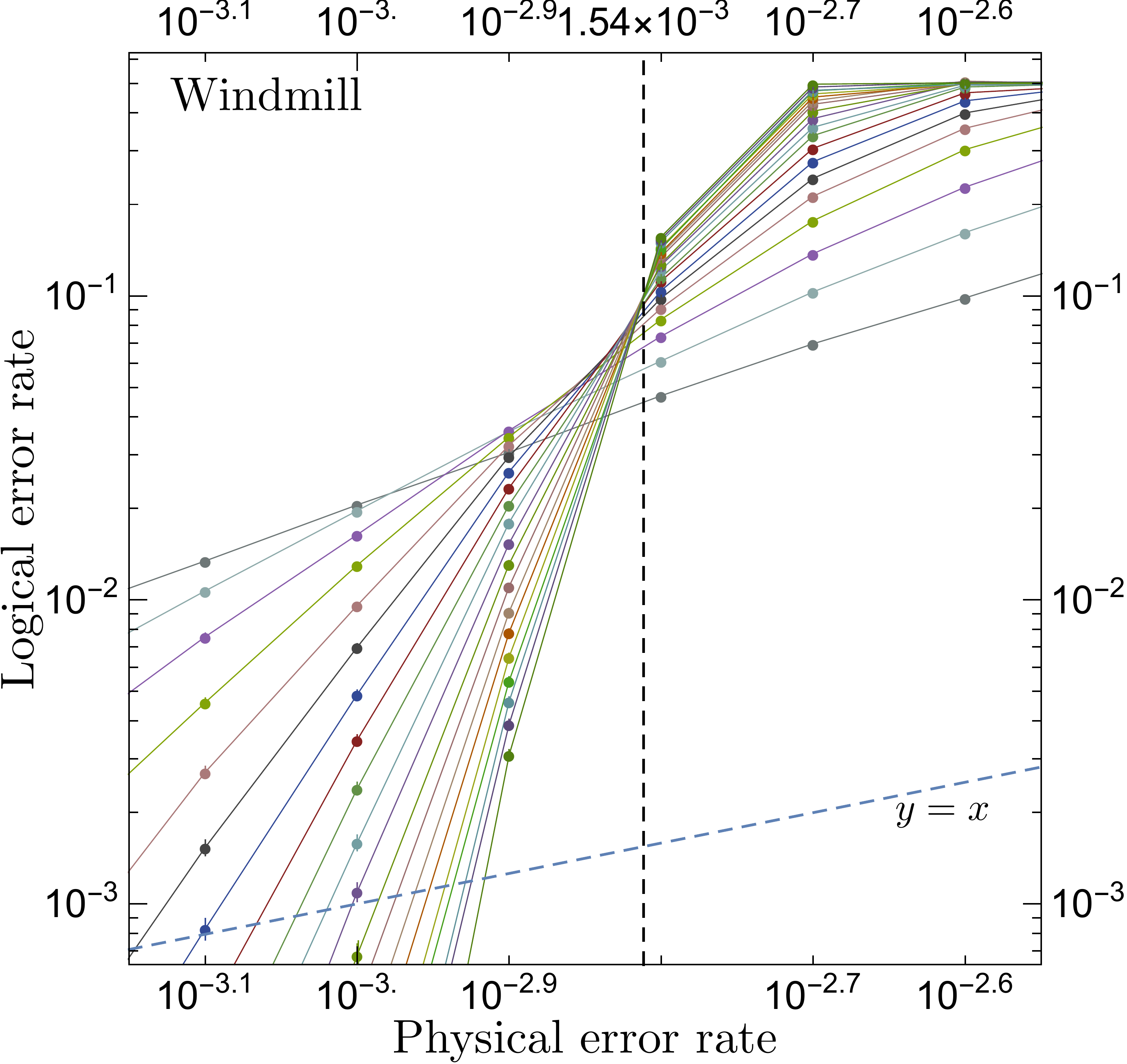}}}
&
\subfigure[\label{fig:d_threshold}]{
\raisebox{.2cm}{\includegraphics[scale=.27]{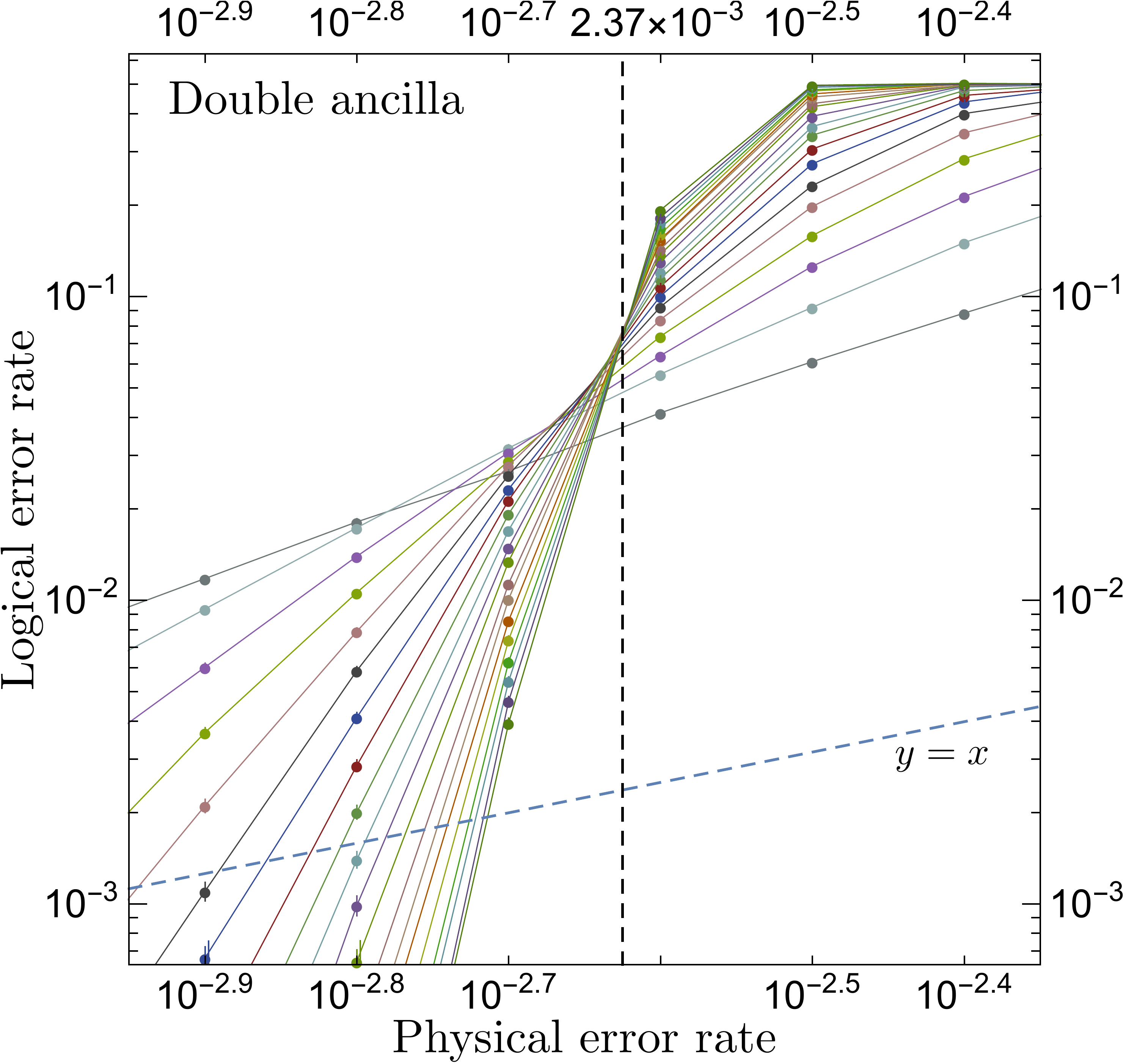}}}
\end{tabular}
\caption{
Logical error rates $p_{\rm L}$ for surface codes with odd distance $d=3,5,\ldots,41$, using the (a) windmill layout as in \figref{fig:w_schedule} and the (b) double ancilla layout as in \figref{fig:d_schedule}.  
Each dot is obtained from $10^6$ trials of Monte Carlo simulation; dots of the same code distance are joined and in same color; error bars indicate 95\% statistical confidence. 
The blue dashed lines $y=x$ manifest the pseudothresholds; the black dashed lines indicate thresholds $p_{\rm{th}} = 1.54\times 10^{-3}$ in (a) and $2.37\times 10^{-3}$ in (b).  
}\label{fig:threshold}
\end{figure*}

\section{Numerical results}
\label{sec:numerics}

\begin{figure*}
\centering
\begin{tabular}{c@{$\quad\quad$}c}
\hspace{-.4cm}\subfigure[\label{fig:w_importance}]{
\raisebox{.2cm}{\includegraphics[scale=.27]{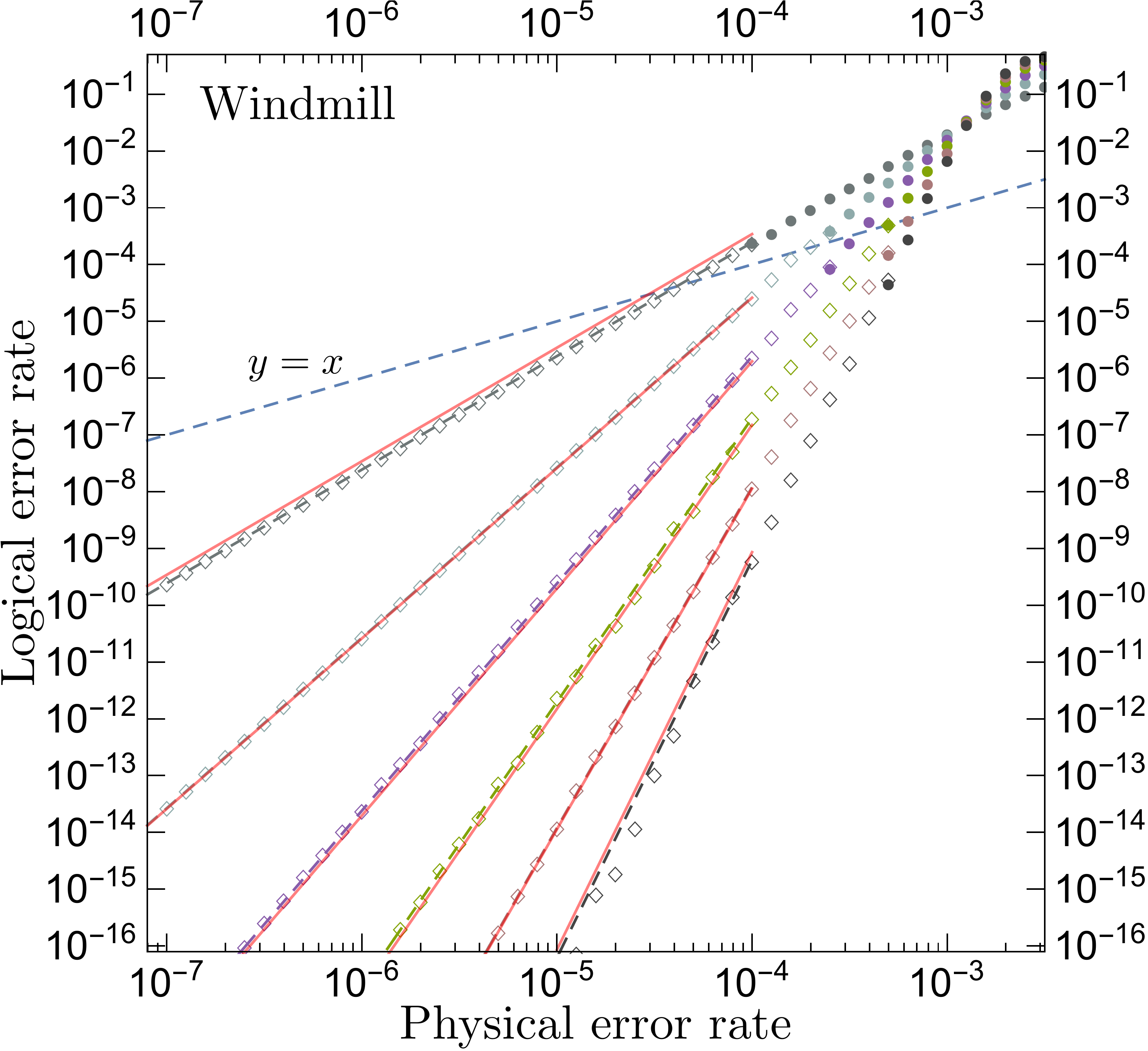}}}
&
\subfigure[\label{fig:d_importance}]{
\raisebox{.2cm}{\includegraphics[scale=.27]{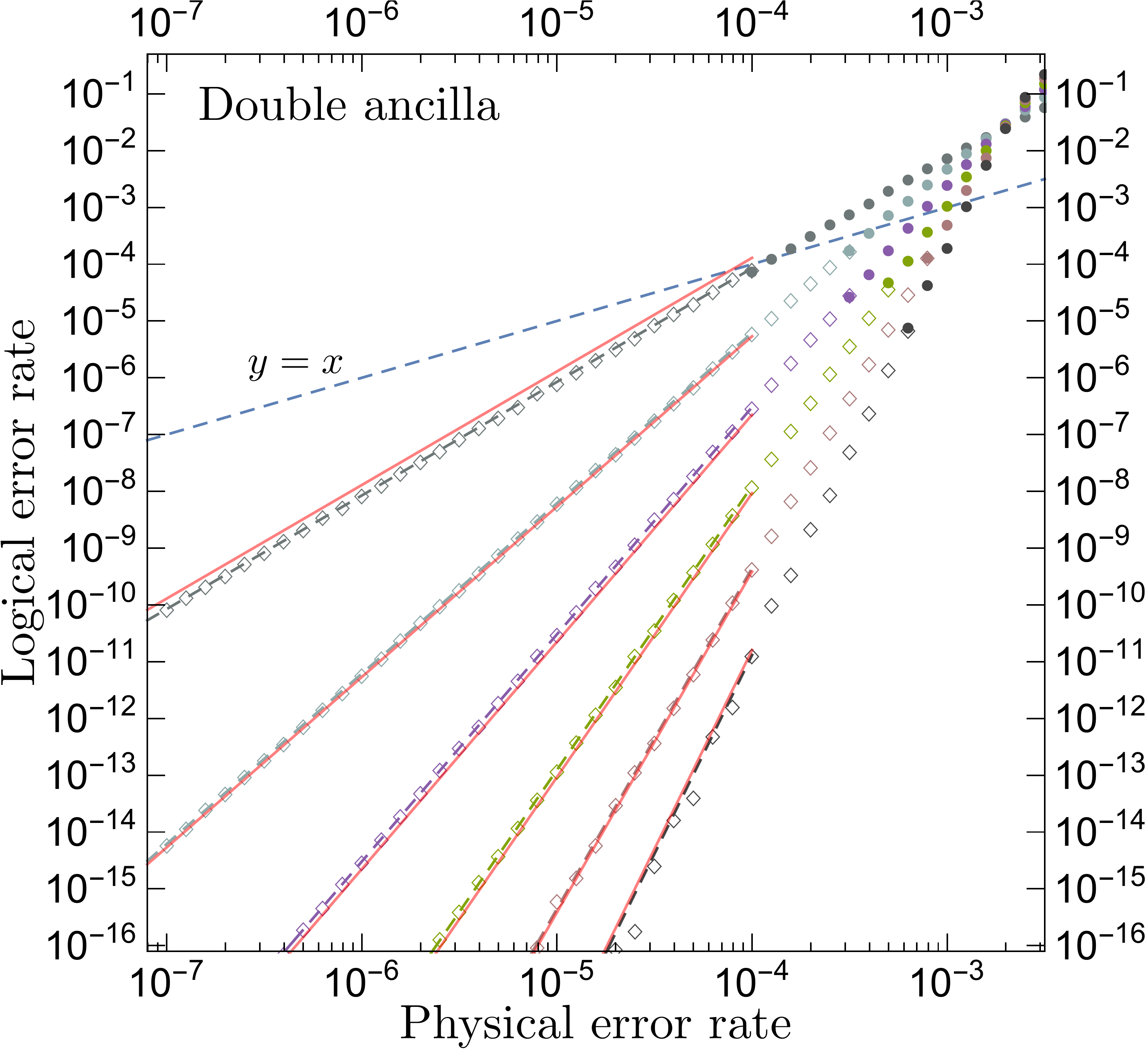}}}
\\[.5cm]
\multicolumn{2}{c}{
\subfigure[\label{tab:importance}]{
\raisebox{3.5cm}{
\begin{tabular}{@{\;\;} @{\;\;} c @{\;\;} @{\;\;} | @{\;\;}  c @{\;\;} @{\;\;} c @{\;\;} c @{\;\;} c @{\;\;}  @{\;\;} c}
\hline \hline \\[-.4cm]
&\multicolumn{2}{c}{Windmill} & & \multicolumn{2}{c}{Double ancilla} 
\\[.06cm]
$d$ & fitted data & $c$ & $\quad$ &  fitted data & $c$ \\[.06cm]
\cmidrule( r){1-1}  \cmidrule(lr){2-3} \cmidrule(lr){5-6} \\[-.4cm]
3 & $[10^{-7}, 10^{-4}]$ & 0.057937 & &
 $[10^{-7}, 10^{-4}]$ & 0.047567 \\
5 & $[10^{-7}, 10^{-4}]$ & 0.097847 & &
 $[10^{-7}, 10^{-4}]$ & 0.082310 \\
7 & $[10^{-7}, 10^{-4}]$ & 0.135503 & &
 $[10^{-7}, 10^{-4}]$ & 0.095642 \\
9 & $[10^{-7}, 10^{-4}]$ & 0.171511 & &
 $[10^{-6}, 10^{-4}]$ & 0.092853 \\
11 & $[10^{-6}, 10^{-4}]$ & 0.159827 & &
 $[10^{-6}, 10^{-4}]$ & 0.079022 \\
13 & $[10^{-5}, 10^{-4}]$ & 0.127366 & &
 $[10^{-5}, 10^{-4}]$ & 0.055062 \\[.1cm]
\cmidrule( r){1-1} \cmidrule(lr){2-3} \cmidrule(lr){5-6} \\[-.5cm]
& all data above &  0.059787  & &
all data above &  0.075034 \\[-.1cm]
& (red lines)  & $(p'_{\textrm{th}} = 0.0013193)$ & &
 (red lines)  & $(p'_{\textrm{th}} = 0.0024074)$ \\ [.1cm]
\hline \hline
\end{tabular}}}}
\end{tabular}
\caption{
Logical error rates $p_{\rm L}$ for distance $d=3,5,7,9,11,13$ in the low-$p$ regime, using the (a) windmill layout as in \figref{fig:w_schedule} and the (b) double ancilla layout as in \figref{fig:d_schedule}. 
Each dot is obtained from $10^6$ trials of Monte Carlo simulation;  diamonds are obtained from importance sampling (see \appref{sec:importance} for details). 
(c) For each layout and $d$, we fit the data points with $p\le10^{-4}$ to model~\eqnref{eqn:cfit}. 
The fitting parameters $c$ are listed in the third and fifth columns, with fitting curves drawn in corresponding colors in the dashed lines in~(a) and~(b). 
We further fit all low-$p$ data with different $d$ to model~\eqnref{eqn:cpfit}. 
The uniform fitting parameters are listed in the bottom row, and the fitting curves are the red lines in (a) and (b). 
}\label{fig:importance}
\end{figure*}

\begin{figure*}
\centering
\begin{tabular}{c@{\!}c@{\!}c}
\hspace{-.7cm}\subfigure[\label{fig:space}]{
\raisebox{.2cm}{\includegraphics[scale=.2]{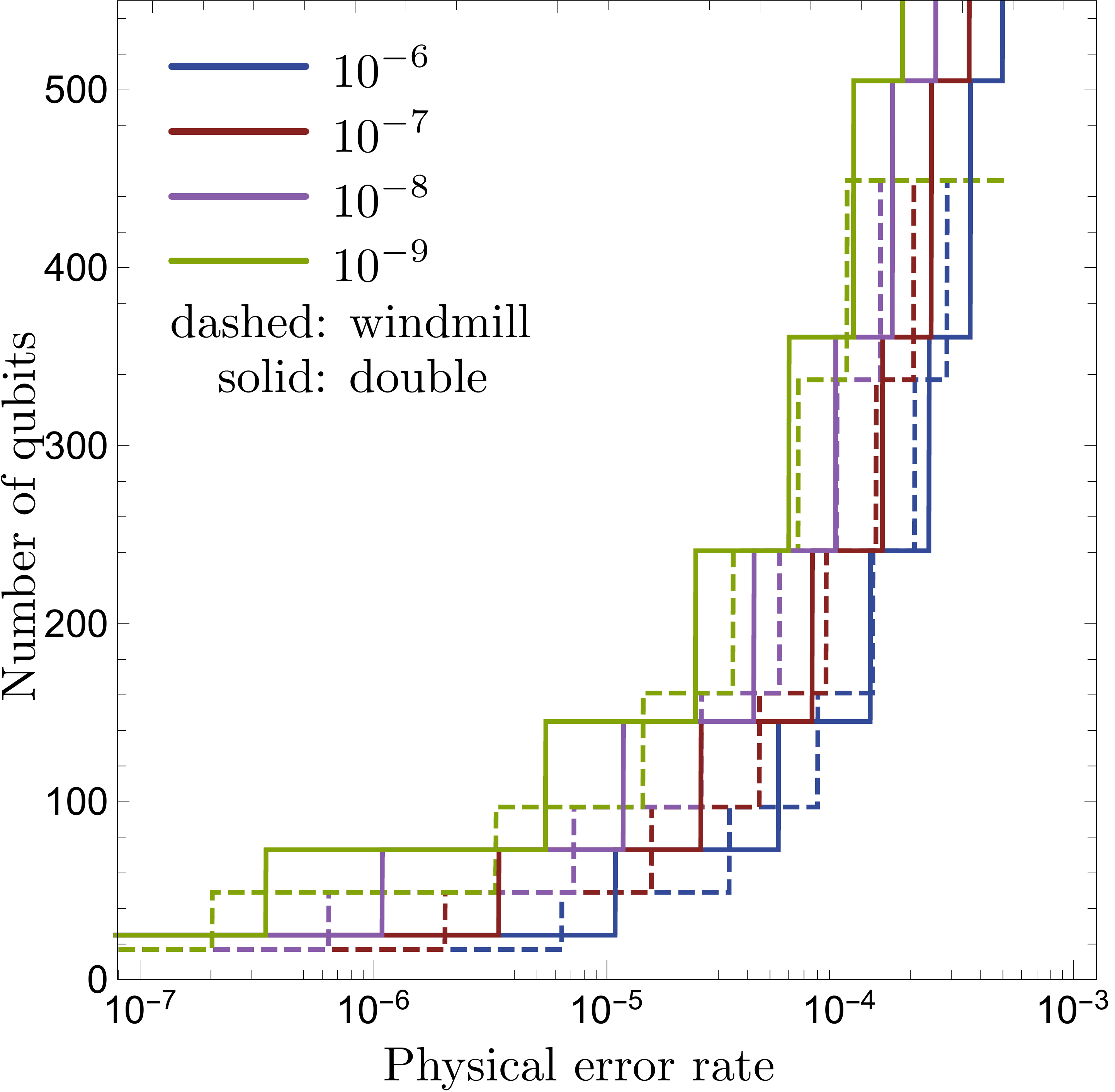}}}
&
\subfigure[\label{fig:time}]{
\raisebox{.2cm}{\includegraphics[scale=.2]{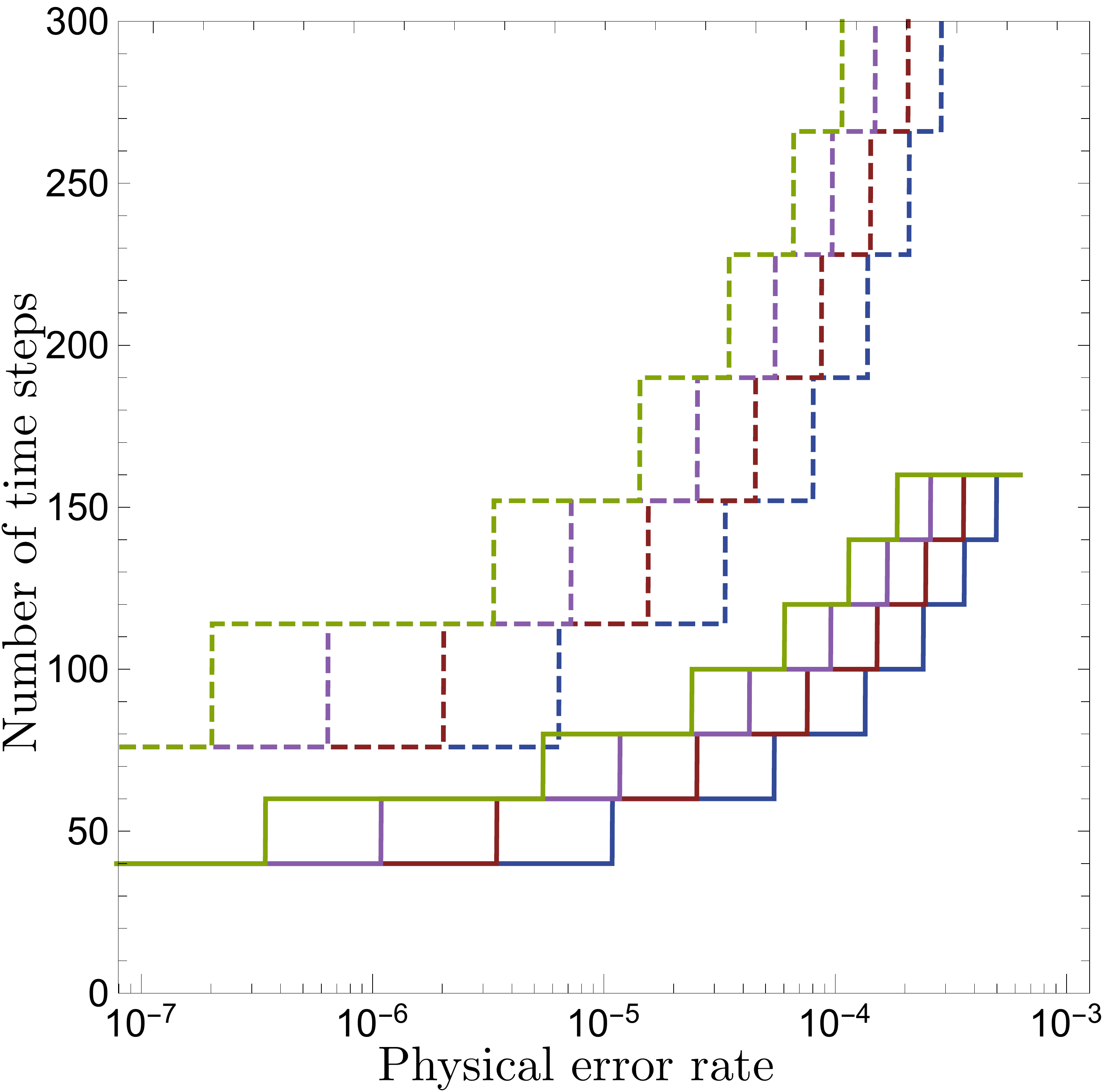}}}
&
\subfigure[\label{fig:spacetime}]{
\raisebox{.2cm}{\includegraphics[scale=.2]{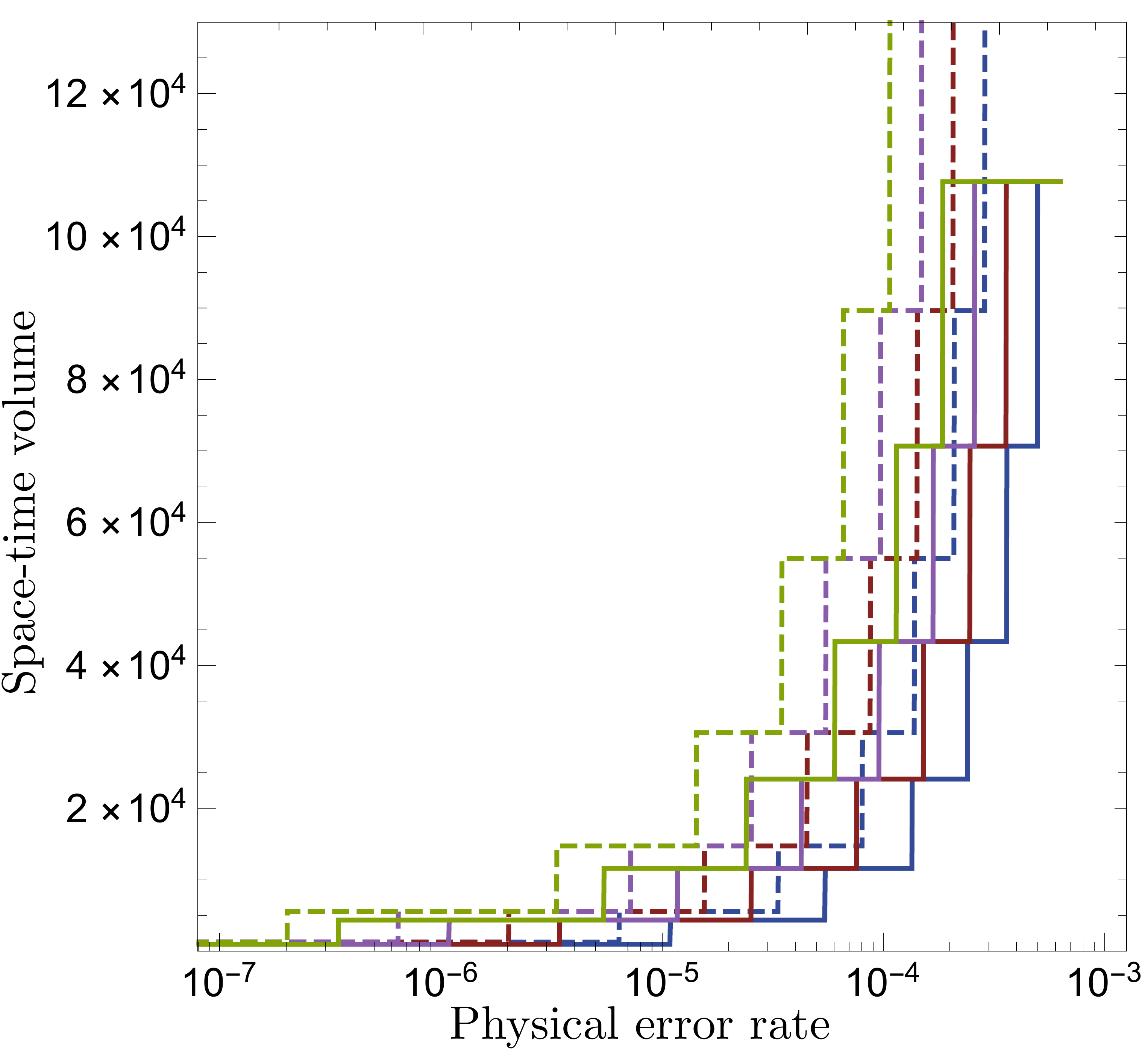}}}
\end{tabular}
\caption{Resource comparison between the windmill layout (dashed) as in \figref{fig:w_schedule} and the double ancilla layout (solid) as in \figref{fig:d_schedule}, in terms of (a) the number of qubits, (b) the number of time steps and (c) the space-time volume, \emph{i.e.}, the product (a) and (b). 
We plot the amount of resources consumed by an error-correction cycle in order to achieve logical error rate $p_\mathrm{L}=10^{-6},10^{-7},10^{-8}$ or $10^{-9}$, under physical error rate $p$.
}\label{fig:overhead}
\end{figure*}

\begin{figure}[!b]
\centering
\begin{tabular}{c}
\subfigure[\label{fig:pseudo}]{
\raisebox{.3cm}{\includegraphics[scale=.25]{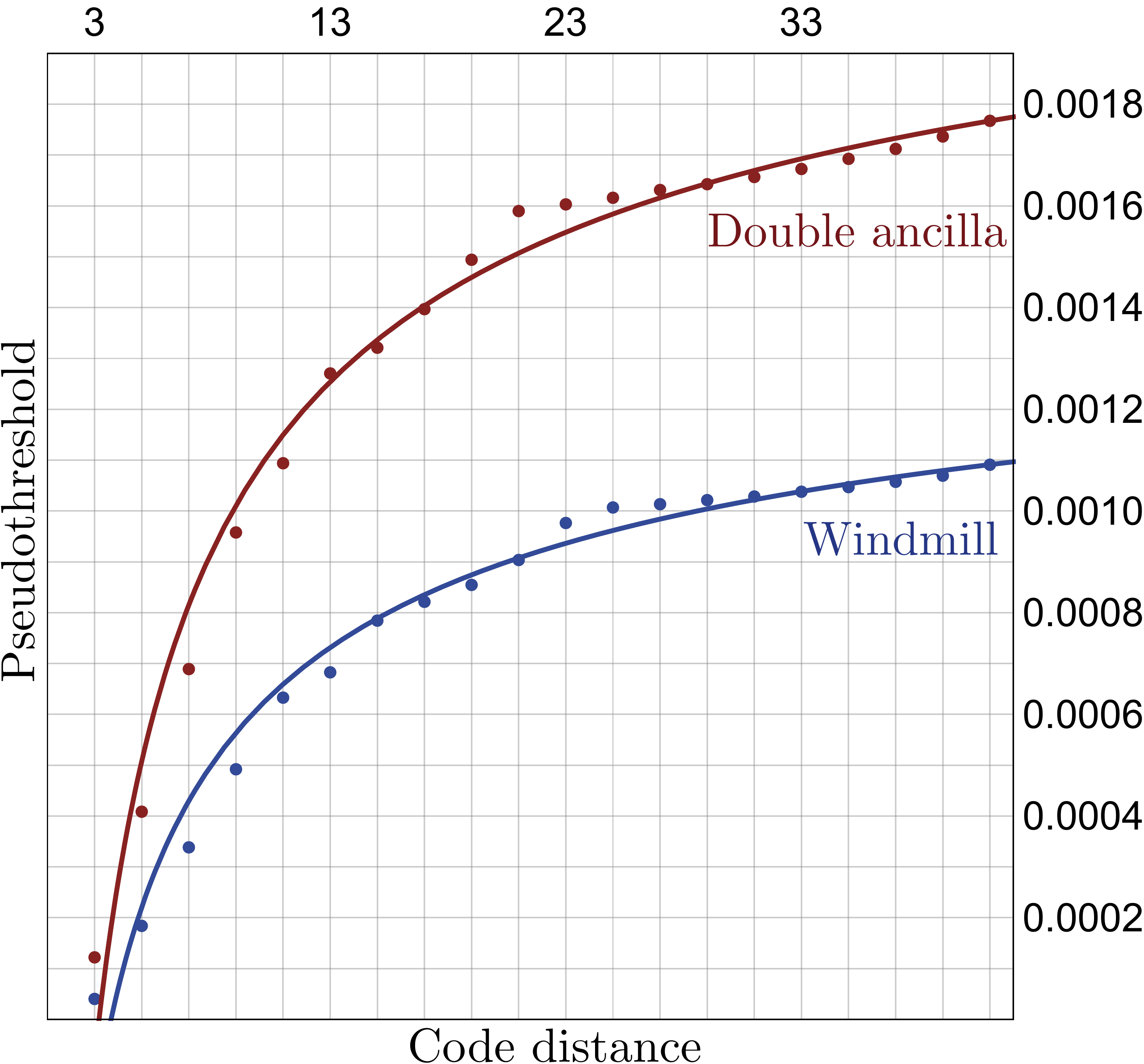}}}
\\
\subfigure[\label{tab:pseudo}]{
\raisebox{3.2cm}{\begin{tabular}{@{\;\;} c @{\;\;} | @{\;\;} c @{\;\;} @{\;\;} c}
\hline \hline \\[-.4cm]
& Windmill & Double ancilla \\
$d$ & $p_{\rm{pseudo}}$ & $p_{\rm{pseudo}}$   \\[.08cm]
\cmidrule( r){1-1}  \cmidrule(lr){2-2} \cmidrule(lr){3-3} \\[-.4cm]
3 &  0.000040202 & 0.000121970 \\
5 &  0.000183853 & 0.000408456 \\
7 & 0.000338334 & 0.000689030\\
9 &  0.000491926 & 0.000957683\\
11 &  0.000632807 & 0.001093850\\[.1cm]
\cmidrule( r){1-1} \cmidrule(lr){2-2} \cmidrule(lr){3-3} \\[-.5cm]& $a=1.95431$ & $a=1.86200$ \\
& $b=0.512408$ & $b=0.536053$ \\ [.1cm]
\hline \hline
\end{tabular}}}
\end{tabular}
\caption{
(a) Dots are pseudothresholds for distance $d=3,5,\ldots,41$, using the windmill layout (blue) as in \figref{fig:w_schedule} and the double ancilla layout (red) as in \figref{fig:d_schedule}.
(b) Values of pseudothreshold for small distance.
Pseudothresholds for all distance up to 41 are fitted using model~\eqnref{eqn:abfit}.
Fitting parameters are in the bottom rows, and the fitting curves are the solid lines in~(a).   
}\label{fig:pseudofit}
\end{figure}

In this section, we present numerical results of the Monte Carlo simulations of our measurement-based surface code error-correction schemes. 
The first in-depth numerical study of the surface code is by Dennis et~al.~\cite{dennis2002surfacecode}, using CNOT-based qubits.
Simulations of the surface code performance and circuit-level optimization have been realized in~\cite{raussendorf2007surfacecode,fowler2009high_threshold,Fowler2010},
providing numerical estimates of the surface code threshold.

\subsection{Methods}
\label{sec:methods}

We estimate the logical error rate with a circuit-level simulation for two layouts: 
the windmill layout as in \figref{fig:windmill} and the double ancilla layout as in \figref{fig:double_ancilla_layout}, 
whose syndrome-extraction circuits are fully specified by \figref{fig:w_schedule} and \figref{fig:d_schedule}, respectively.  
Weight-four plaquettes in the bulk are measured using the gadgets in \figref{fig:compressedX4}, 
with the joint measurements on data qubits scheduled in \texttt{Z} or \texttt{N} order. 
Weight-two plaquettes on the boundaries are measured using the gadgets in \figref{fig:compressedX2}.
For the windmill layout, $X$- and $Z$-type stabilizers are measured in separate stages, 
whereas for the double ancilla layout all stabilizers are measured simultaneously. 

Given a specific syndrome-extraction circuit, we simulate the error process  according to the model as in \secref{sec:noisemodel}, 
and calculate corrections using the Union-Find decoder.
Below we remark on the numerical methods of detecting logical errors.

\smallskip 

In Ref.~\cite{fowler2009high_threshold} the surface code of distance $d$ was considered with time boundary conditions
such that initially the physical qubits are in a code state without error,
syndrome bits are extracted for $T$ rounds by a noisy circuit,
and then a final round of syndrome bits are obtained with a \emph{noiseless} circuit.
The value of $T$ is increased until a logical error is observed.
This setting makes it easy to detect any logical error introduced by errors and their correction,
but we were unable to find an operational meaning to their time boundary conditions.
In particular, the presence of the noiseless syndrome measurement at the end
could result in the underestimation of the logical error rate because, in principle, a decoder may exploit the information from the last perfect syndrome measurement.
A potential justification for this could be that ultimately (for instance at the end of a quantum computation) each qubit in the surface code will be measured out qubit by qubit,
allowing for a more reliable syndrome readout than usual.
However, while modeling the performance of the surface code far from this final readout, it seems important to ensure that the model is not sensitive to this artificial step.

We have observed that the time boundary conditions make the logical error rate underestimated
by a factor of~$\approx 1.5$, independent of the code distance,
when the physical error rate is $10^{-3}$. 
See \appref{app:timeboundary}.
Based on this observation,
we continue to use the same time boundary conditions as in~\cite{fowler2009high_threshold}
with~$d$ rounds of noisy syndrome measurement.
That is, we start with a code state, perform~$d$ rounds of noisy syndrome measurements,
and then finish by one round of noiseless syndrome measurement.
The whole history is passed to the Union-Find decoder,
and we define the storage error rate $p_{\rm L}$ 
to be the probability that this procedure results in a nontrivial logical operator.

We believe that in future work it is desirable to have more operationally meaningful estimations of the failure rate of logical operations; these should not be limited to the examination of identity gates but include embedding a patch into a larger one (growing), logical multi-qubit Pauli measurements, and logical single-patch measurement after idling.

\smallskip

In \appref{sec:unionfind}, we argue that the Union-Find decoder succeeds in error correction 
as long as there are at most $\left\lfloor \tfrac{d-1}{2}\right\rfloor$ faults in a single trial with $d$ noisy rounds as outlined above,  
thereby maintaining an effective distance of the surface code.


Throughout our simulations, instead of randomly generating faults and tracing them through the circuit to determine a history of syndromes,
we obtain the syndromes straightforwardly by sampling edges on a decoding graph. 
This edge sampling method has been explicitly used to obtain simulation speedups by Newman {\em et al.}~\cite{newman2020generating}, who assume error models other than the depolarizing model as considered here.
In \appref{app:inclusivemodel}, we establish the efficacy of the edge sampling method for the depolarizing noise model by mapping it to an equivalent {\em inclusive error model}, which enables accelarated yet faithful simulations.

\subsection{Results}

Figure~\ref{fig:threshold} plots the logical error rate $p_{\rm{L}}$ of our error-correction schemes for surface codes with odd distance $d=3,5,\ldots,41$ with relatively high physical error rate~$p$.  
Each dot is obtained from $10^6$ trials of Monte Carlo simulation; error bars indicate 95\% statistical confidence. 
We observe empirical thresholds $p_{\rm{th}} = 1.54\times 10^{-3}$ for the windmill layout and $2.37\times 10^{-3}$ for the double ancilla layout. 

Figures~\ref{fig:w_importance} and~\ref{fig:d_importance} plot $p_{\rm{L}}$ for surface codes with distance $d=3,5,7,9,11,13$ in the low-$p$ regime. 
Each dot is obtained from $10^6$ trials of Monte Carlo simulation; diamonds are obtained from importance sampling.  
We explain importance sampling in details in \appref{sec:importance}. 
Following the heuristic in~\cite{fowler2012surfacecode}, for each $d$, we fit the data points in the relatively low error regime (with $p\le 10^{-4}$) to the model 
\begin{equation}\label{eqn:cfit}
p_{\rm{L}} = c(d) \cdot (p/p_{\rm{th}}) ^ {(d+1)/2}, 
\end{equation}
where $c(d)$ is a constant that only depends on $d$, and $p_{\rm{th}}=1.54\times 10^{-3}$. 
Since our schemes can correct up to $\left\lfloor \tfrac{d-1}{2} \right\rfloor$ faults, provided $p$ is small~\eqnref{eqn:cfit} should be a reasonable heuristic. 
The fitting parameters $c(d)$ are listed in \tabref{tab:importance},  and the fitting curves are depicted in corresponding colors in dashed lines in \figref{fig:w_importance} and \figref{fig:d_importance}.
Since the values of $c(d)$'s are  comparable to each other, we continue to fit all the low-$p$ regime data with different $d$ to a uniform heuristic 
\begin{equation}\label{eqn:cpfit}
p_{\rm{L}} = c \cdot \big( p/p'_{\rm{th}}  \big)^{(d+1)/2},
\end{equation}
where $c$ and $p'_{\rm{th}}$ are both to be fitted, independent of~$d$. 
The fitting parameters $c$ and $p'_{\rm{th}}$ are given in the bottom row of \tabref{tab:importance}, and the fitting curves are depicted by red lines in \figref{fig:w_importance} and \figref{fig:d_importance}.

Figure~\ref{fig:overhead} compares the resources consumed by schemes based on the windmill layout and the double ancilla layout.
Using the fitting parameters in \figref{fig:importance}, we calculate (a) the number of qubits, (b) the number of time steps and (c) the space-time volume, \emph{i.e.}, the product of (a) and (b), which are required by an error-correction cycle in order to achieve a logical error rate of $10^{-6},10^{-7},10^{-8}$ or $10^{-9}$.
It can be observed that, compared with that based on the windmill layout, the scheme based on the  double ancilla layout consumes more qubits for most physical error rates, while always requires less time and space-time volume.

Figure~\ref{fig:pseudo} plots the pseudothresholds for $d=3,5,\ldots,41$ by dots, and fits them with the solid curves, using the following heuristic
\begin{equation}\label{eqn:abfit}
p_{\rm{pseudo}} = p_{\rm{th}} \cdot (1- a \cdot d^{-b}),
\end{equation}
where $p_{\rm{th}}=1.54\times 10^{-3}$ or $2.37\times10^{-3}$, and $a,b$ are to be fitted, independent of $d$. 
We choose the heuristic~\eqnref{eqn:abfit} because we find that the relation between $\log (p_{\textrm{th}} - p_{\textrm{pseudo}})$ and $\log d$ is close to linear. 
Table~\ref{tab:pseudo} lists the values of pseudothreshold for small distance, along with the fitting parameters for all distance up to 41. 

\smallskip

Observe that in general the scheme using the double ancilla layout has lower logical error rate and higher threshold or pseudothreshold than the one using the windmill layout.
This better performance is consistent with the fact that the space-time volume of the double ancilla layout is smaller than that of the windmill layout. 

\section{Conclusion}

We have described several surface code error-correction schemes that are  tailored for measurement-based qubits, {\em i.e.}, hardware equipped with Pauli measurements on single qubits and pairs of nearest-neighbor qubits. 
Instead of directly translating from the canonical CNOT-based scheme, our schemes feature a hardware-efficient qubit layout and an optimized syndrome-extraction circuit, together giving rise to reasonable error thresholds.  
We have also designed alternative surface code layouts and measurement circuits for general Pauli operators, which might be of independent interest. 

It remains to develop systematic methods for constructing more efficient measurement-based syndrome-extraction circuits. 
More work has to be done to investigate the tradeoff among the circuit depth, qubit connectivity and ancilla overhead.  

\begin{acknowledgements}
This paper is dedicated to David Poulin, a friend, a mentor, and an inspiration to the quantum computing community.
R.~C. thanks Microsoft Quantum for hospitality during his internship; Qian Yu for helpful discussions; NSF grant CCF-1254119, ARO grant W911NF-12-1-0541 and MURI Grant FA9550-18-1-0161 for partial support. 
\end{acknowledgements}

\begin{figure*}[t]
\centering
\begin{tabular}{c}
\subfigure[\label{fig:compressedX2}]{
\raisebox{.1cm}{\includegraphics[scale=.8]{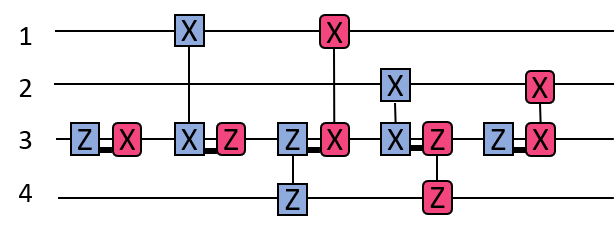}}}
\\
\hspace{-.5cm}\subfigure[\label{fig:compressedXeven}]{
\raisebox{.5cm}{\includegraphics[scale=.55]{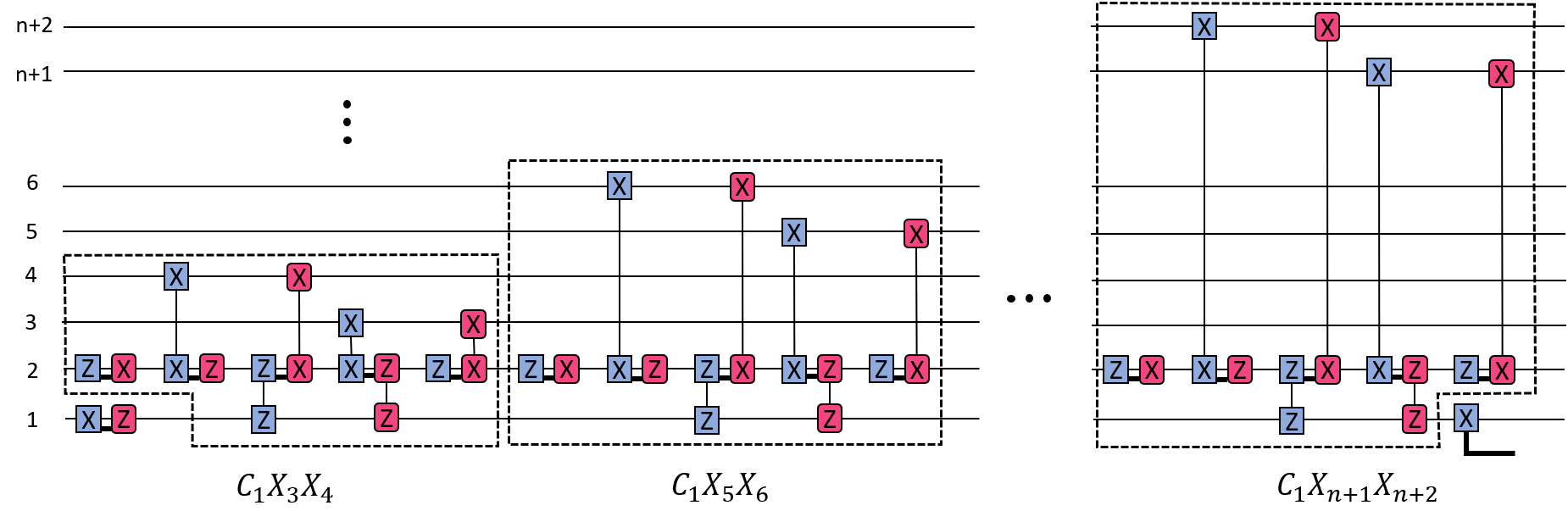}}}
\end{tabular}
\caption{
(a) A circuit which implements a 2-target controlled-NOT gate, with qubit 4 as the control, qubits 1 and 2 as the targets, and where qubit 3 is an ancilla.
(b) The 2-target controlled-NOT gate can be bootstrapped to implement an arbitrary weight-$n$ measurement, where $n$ is even. 
The dangling thick line represents that the given measurement outcome encodes the overall measurement outcome of $X^{\otimes n}$.
}\label{fig:Xeven}
\end{figure*}

\begin{figure*}[t]
\centering
\begin{tabular}{c}
\subfigure[\label{fig:compressedX3}]{
\raisebox{.5cm}{\includegraphics[scale=.8]{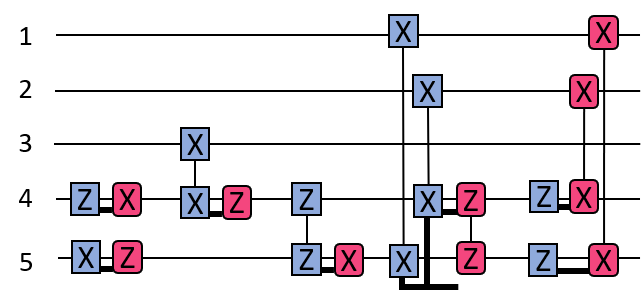}}}
\\
\hspace{-.5cm}\subfigure[\label{fig:compressedXodd}]{
\raisebox{.5cm}{\includegraphics[scale=.55]{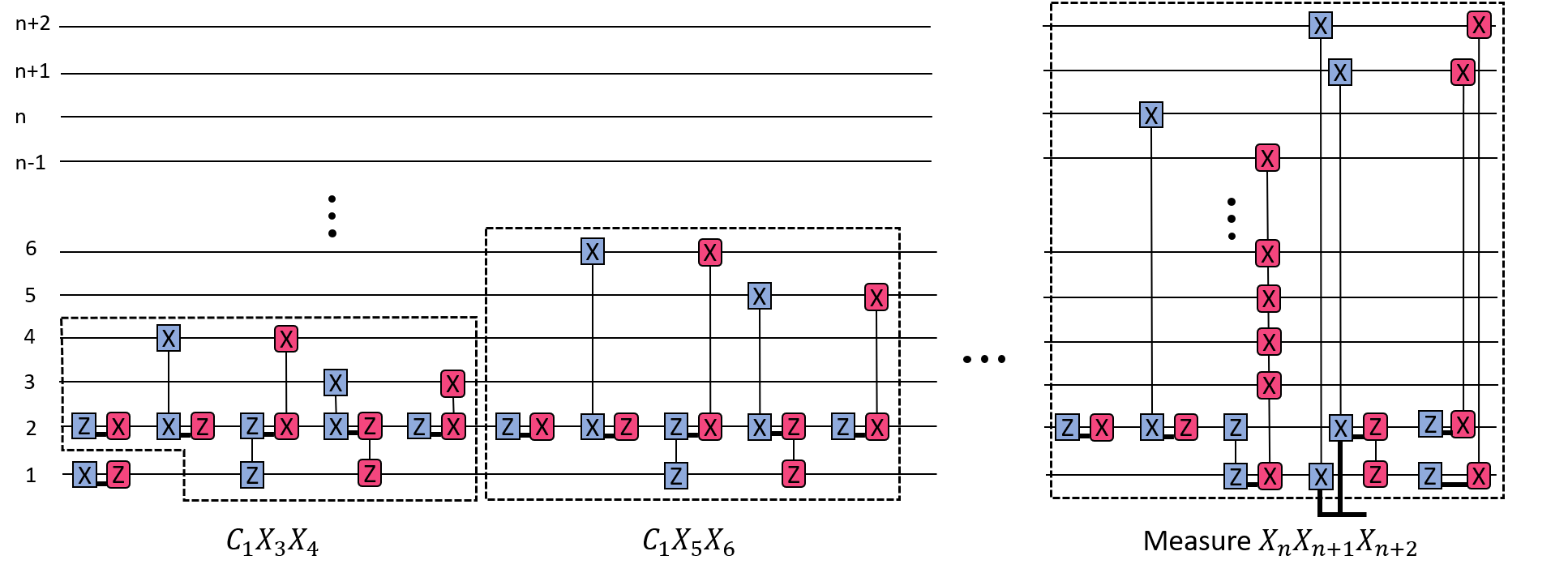}}}
\end{tabular}
\caption{
(a) A circuit which implements the $X^{\otimes 3}$ measurement on qubits 1, 2 and 3, using qubits 4 and 5 as ancillas.
This is more efficient than using a 2-target controlled-NOT gate followed by a single CNOT gate.
Note that the measurement outcome of $X^{\otimes 3}$ is obtained from the parity of those of a pair of measurements, indicated by a dangling  junction of thick lines.
(b)~By adding this (slightly modified) to the end of a sequence of 2-target controlled-NOT gates, we obtain a general scheme for measuring $X^{\otimes n}$ for odd~$n$.
Notice the first $X$ measurement on the bottom ancilla qubit is omitted in going from (a) to (b). 
Also note that the $Z^{\otimes 2}$ measurement on the two ancilla qubits 1 and 2 in the $X_nX_{n+1}X_{n+2}$ block has a Pauli update which is supported on all but the top three data qubits.
}\label{fig:Xodd}
\end{figure*}

\appendix

\section{Optimizing general Pauli measurement circuits}
\label{sec:optimizing-general-measurements}

Here we consider optimizations of the circuit built from single-qubit and joint measurements to measure a general $n$-qubit Pauli operator $P$.
These can be used to measure the stabilizer generators of any stabilizer code, including LDPC codes, surface codes, color codes etc., and also the gauge generators of any Pauli subsystem code. 

Our goal will be to minimize the number of ancilla qubits and measurements which are required.
We assume here that any single-qubit Pauli measurement is possible on any qubit, and that any joint Pauli measurement is possible on any pair of qubits.
First we note that any circuit which measures a Pauli $P$ is equivalent to a circuit to measure $X^{\otimes n}$ since one can move between the two circuits using single-qubit Clifford operations.
Therefore we will focus on measuring $X^{\otimes n}$, but this can be straightforwardly converted to a circuit for measuring any other weight-$n$ Pauli operator with the same number of ancilla qubits, connectivity and single-qubit and joint measurements (albeit in  different measurement bases).

Our general approach is to split the $n$ relevant qubits into subsets $n=\sum_i m_i$, then to prepare an ancilla in $\ket{+}$, and sequentially apply $m_i$-target controlled-NOT gates from that ancilla to the subsets of $m_i$ qubits, before finally measuring the ancilla in the $X$ basis to read off the measurement outcome. 
Then we can separately optimize the modular component of each $m_i$-target controlled-NOT gate.
The trivial case is where $m_i=1$ for all $i$, and we therefore break the measurement up into a sequence of $n$ CNOT gates each implemented as in \figref{fig:cnot}.
This would require 2 ancilla qubits, and $4n+2$ measurements ($2n+2$ single-qubit measurements and $2n$ joint measurements). 

We now focus on $m_i=2$, {\em i.e.}, optimizing the 2-target controlled-NOT $\mathrm{C}X\!X$ gate.
By numerically searching over measurement-based circuits we have found the circuit shown in \figref{fig:compressedX2}.
When $n$ is even, we can use this approach to construct a circuit which measures $X^{\otimes n}$ and uses two ancilla qubits, $5n/2+2$ measurements ($n+2$ single-qubit measurements, and $3n/2$ joint measurements) as shown in \figref{fig:compressedXeven}.
Also note that when $n=4$ this recovers the circuit described in \figref{fig:compressedX4}, which we use for the implementation of the surface code with measurement-based qubits.

Further reduction in the number of measurements is possible; for example,  note that in \figref{fig:compressedXeven} $Z$ is measured at the end of $\mathrm{C}_1X_3X_4$, and then again immediately after at the beginning of $\mathrm C_1X_5X_6$. 
One of these can clearly be removed; however, it is worth noting that this removal affects how errors propagate within the circuit, and may result in a less robust measurement of $X^{\otimes n}$ with regard to faults.

Suppose now that $n$ is odd. 
The most naive strategy is to use the circuit obtained from using $n$ CNOT gates each implemented as in \figref{fig:cnot}, and would require 2 ancilla qubits, and $4n+2$ measurements ($2n+2$ single-qubit measurements and $2n$ joint measurements). 
A better strategy is to use what we have found above to implement $(n-1)/2$ 2-target controlled-NOT $\mathrm CX\!X$ gates followed by a single CNOT gate.
This would require 2 ancilla qubits, and $(5n+9)/2$ measurements ($n+4$ single-qubit measurements, and $(3n+1)/2$ joint measurements). 
However, there is yet a better way---consider the gadget shown in \figref{fig:compressedX3} to measure $X^{\otimes 3}$ directly.
We can use a sligthly modified version of this $X^{\otimes 3}$ measurement circuit in combination with the circuit in \figref{fig:compressedX2} $(n-3)/2$ times for the remaining $n-3$ qubits in the measurement; see \figref{fig:compressedXodd}.
This requires 2 ancilla qubits and $(5n-3)/2$ measurmements ($n$ single-qubit measurements and $3(n-1)/2$ joint measurements).

We also include an efficient implementation of the swap circuit in \appref{sec:swap-appendix}, which might be of independent interest.

\section{Efficient measurement-based swap circuit}
\label{sec:swap-appendix}

\begin{figure}[t]
\centering
\includegraphics[scale=.8]{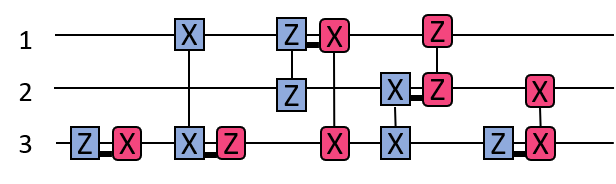}
\caption{
An optimized version of the swap circuit, which swaps qubits 1 and 2 using qubit 3 as an ancilla. 
}\label{fig:swap_circuit}
\end{figure}

In \figref{fig:swap_circuit}, we show a swap circuit which uses one ancilla qubit, and 5 measurements (2 single-qubit measurements and 3 joint measurements). 
The naive implementation is built from 3 CNOT gates as in \figref{fig:cnot} and requires 12 measurements (6 single-qubit measurements and 6 joint measurements).

\section{Time boundary conditions for logical error rate estimation}
\label{app:timeboundary}

\begin{figure}[t]
\includegraphics[width=0.45\textwidth]{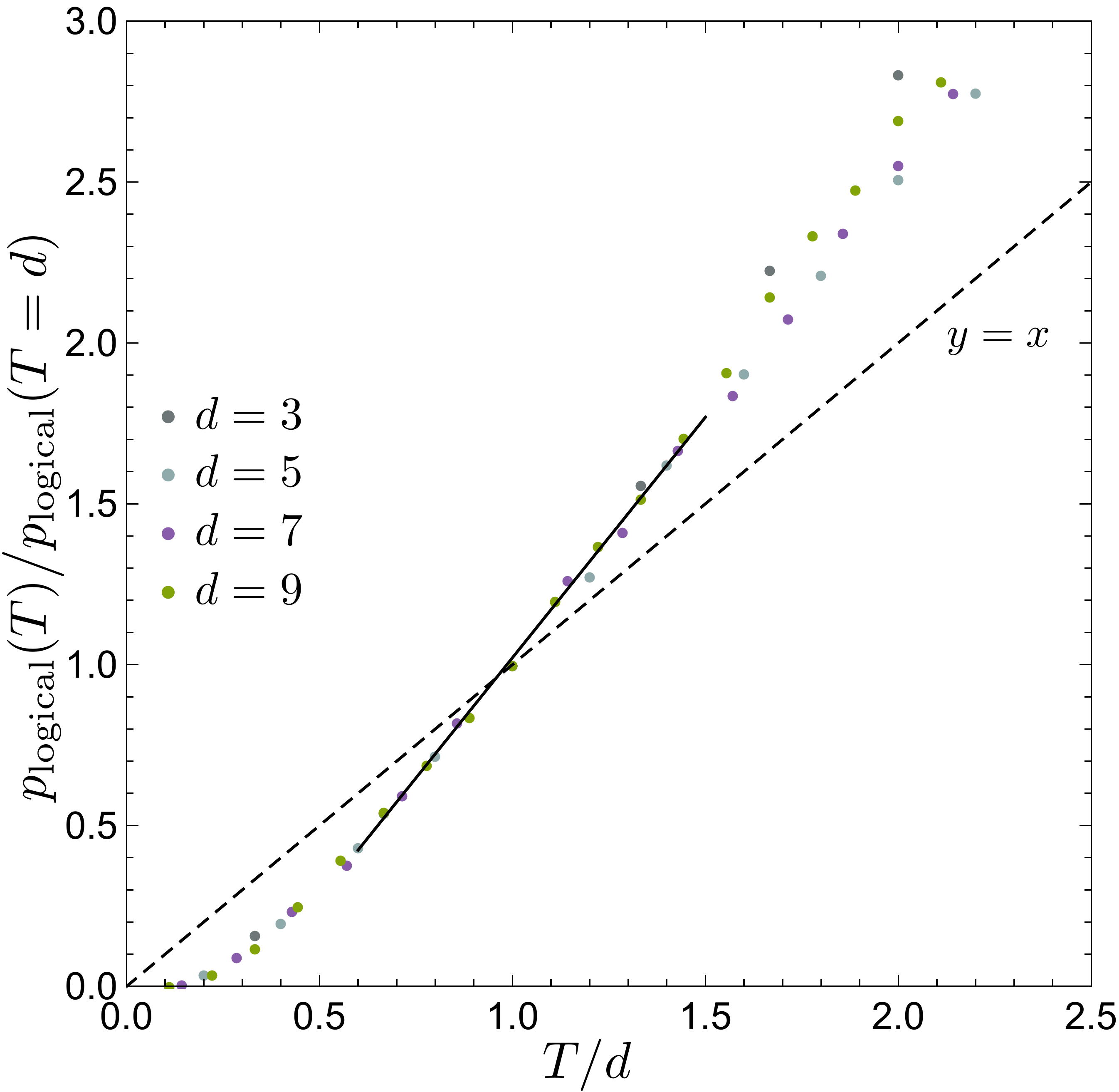}
\caption{Logical error rates within varying time windows 
for distance $d=3,5,7,9$ with physical error rate $p=10^{-3}$, using the double ancilla layout as in \figref{fig:d_schedule}.  
Each dot is obtained from $10^6$ trials of Monte Carlo simulation; individual trials start with a code state with no error, extract syndromes with noisy circuits for $T=1,2,\ldots,20$ rounds and end with a noiseless measurement  round.     
Rescaled dots with $0.6\le T/d \le 1.5$ collapse to the black solid line, suggesting~\eqnref{eq:linearapprox} with $\alpha \approx 1.5$.  
See \appref{sec:unionfind} and \appref{app:inclusivemodel} for more simulation details.  
}\label{fig:varyingTimeDimension}
\end{figure}

We continue the discussion on the issue of time boundary conditions,
started in \secref{sec:methods},
to measure logical error rate of the surface code using noisy circuits.

Ideally, we would measure the probability $p_{\rm{logical}}^{\rm{ideal}}$ of the event 
that errors and the correction operator together form a nontrivial logical operator
in a given unit time window, assuming that the memory has existed and will exist for an indefinite period of time.
This practically irrelevant but mathematically sound scenario, poses a problem to numerics
since no decoding algorithm can take the infinite history of syndrome measurements.
However, since errors and the corresponding correction operator 
have exponentially decaying correlation in time given a reasonable decoding algorithm~\cite{dennis2002surfacecode},
it should suffice to consider a finite time segment to measure $p_{\rm{logical}}^{\rm{ideal}}$.

Let $p_{\rm logical}(T)$ be the probability 
that there is a logical error in the setting of~\cite{fowler2009high_threshold}
when there are $T$ rounds of noisy and one additional round of noiseless syndrome measurement.
Since errors and correcting operators have short time correlations with the Union-Find decoder (and with the minimum weight matching decoder),
we may expect that $p_{\rm{logical}}$ is a reasonable proxy to $p_{\rm{logical}}^{\rm{ideal}}$ in our setting.
We believe that $p_{\rm{logical}}(T)$, as a function of~$T$,
converges for large $T$ to a linear function 
\begin{align}
p_{\rm{logical}}(T) \sim \alpha \, T + \beta \label{eq:linearapprox}
\end{align}
for any fixed physical error rate $p$ and code distance~$d$.
Then, $p_{\rm{logical}}^{\rm{ideal}}$ can be identified with the coefficient $\alpha$ 
times the unit memory time which can be $d$:
\begin{align*}
p_{\rm{logical}}^{\rm{ideal}} = \alpha \, d.
\end{align*}
We have confirmed \eqnref{eq:linearapprox} for $p=10^{-3}$ (physical error rate) 
with the double ancilla layout as depicted in \figref{fig:double_ancilla_layout}; 
see \figref{fig:varyingTimeDimension}.
We observe that
\begin{align*}
p_{\rm{logical}}^{\rm{ideal}} \approx 1.5 \, p_{\rm{logical}}(T=d)
\end{align*}
independent of $d$ for $p = 10^{-3}$.

All the logical error rates we report in this paper use
\begin{align*}
p_{\rm{L}} : = p_{\rm{logical}}(T=d)
\end{align*}
as the storage error rate of our surface code.

\section{Union-Find decoder}
\label{sec:unionfind}

Here we briefly explain how to use the Union-Find decoder~\cite{delfosse2017unionfind, delfosse2017peeling} to correct errors in the surface code given a fixed qubit layout and syndrome-extraction circuit. 


As explained in \secref{sec:methods} and \appref{app:timeboundary}, in a single trial of Monte Carlo simulation for a distance-$d$ surface code, we start with all the qubits without error, and then repeat the syndrome extraction with faulty circuits for $d$ rounds, followed by an additional round with noiseless circuit.  

Among all the popular surface code decoders, we choose to use the Union-Find decoder due to its simplicity and rapidity.
As is typical with CSS codes, $X$-type and $Z$-type errors on the surface code can be dealt with separately. 
For simplicity, we only care about $X$ stabilizer syndromes throughout the simulations, and evaluate the logical error rate $p_{\rm L}$ as the probability of having a logical $Z$ error, whereas the circuit used in the simulation still extracts both $X$ and $Z$ stabilizer syndromes. 
We consider the simplest version of the Union-Find decoder without weighted growth 
which grows small clusters first~\cite{delfosse2017unionfind}.
We have not tried to improve the Union-Find decoder by exploiting the correlation between the two types of syndromes~\cite{fowler2013correlated, delfosse2014correlations}. 
We have also not tried the recent optimized version of the Union-Find decoder~\cite{huang2020fault, huang2020weightedUF}, which might lead to better performance at the price of a slightly more complex decoding algorithm.

\begin{figure}
\centering
\includegraphics[scale=.22]{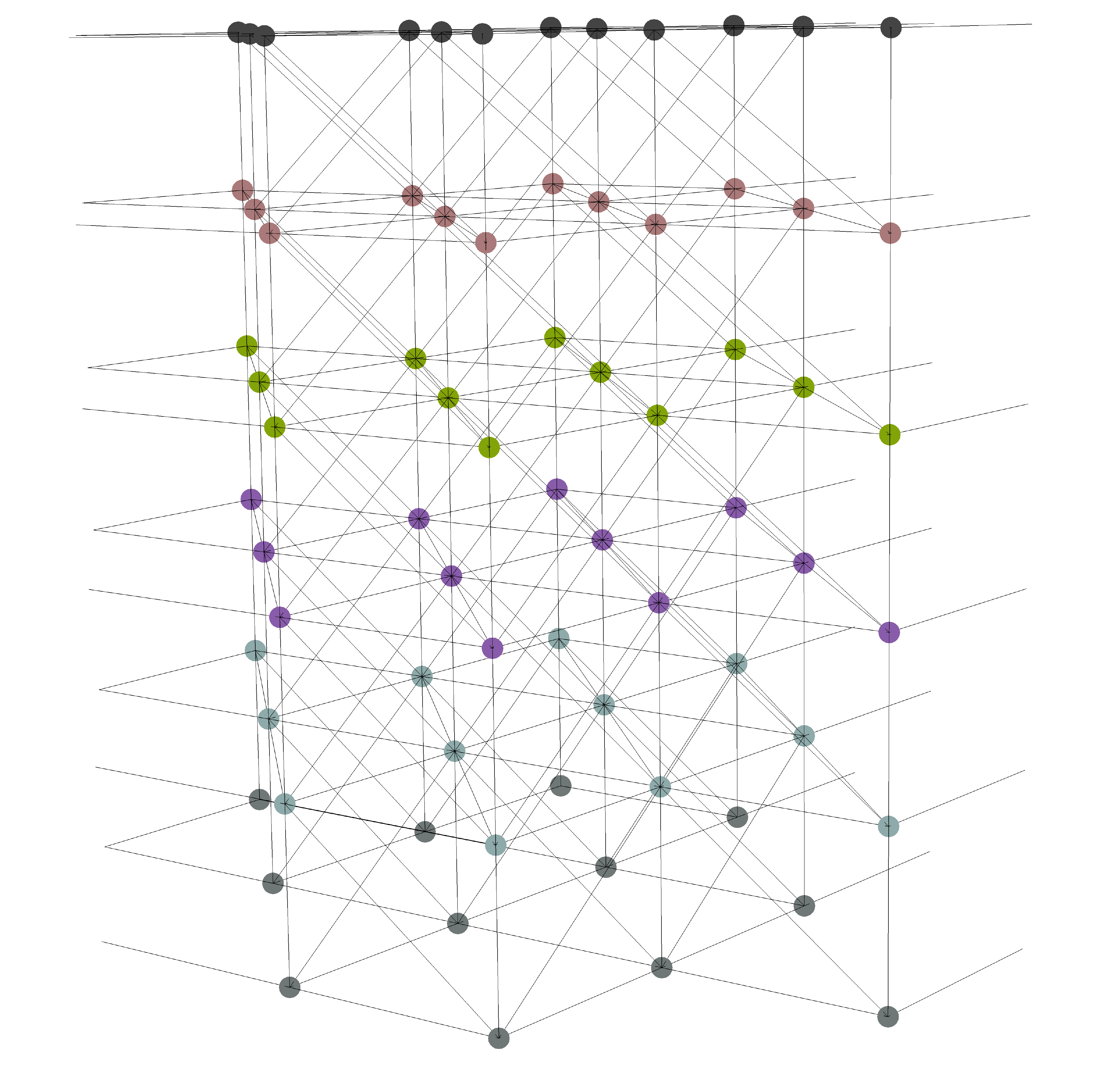}
\caption{
The decoding graph for the distance-five surface code  using the windmill layout and optimized syndrome extraction circuit as specified in \figref{fig:w_schedule}. 
Time proceeds up vertically. 
The vertices within each of the six layers (in the same color) correspond to the changes between the $X$ syndrome bits extracted in that round and those in the previous round. 
All the dangling edges on the space boundaries are connected to a same vertex $b$ (not depicted). 
Edges exist between those vertex pairs whose triviality are changed by a single fault.
The decoding graph using the circuit as in \figref{fig:d_schedule} with the double ancilla layout is similar. 
}\label{fig:decoding_graph}
\end{figure}

A useful way to analyze the decoding algorithm is  to imagine the space-time error-correction circuit as a three-dimensional \emph{decoding graph} $G=(V,E)$; see \figref{fig:decoding_graph}.
Specifically, $V=\{b\}\cup\left(\bigcup_{\tau =1}^{d+1}V_\tau\right)$,
where $V_\tau$ are all identical to one another as sets.
This reflects the fact that we repeatedly measure (via our noisy circuit) the same set of stabilizers.
A syndrome bit measured in round $\tau = 1,2,\ldots, d$ 
corresponds to a vertical edge that connects vertices, one in $V_\tau$ and the other in $V_{\tau+1}$.
Given $d+1$ rounds of observed syndrome bits, 
we call a vertex $v \in V_\tau$, $\tau=1,\ldots,d+1$ to be \emph{nontrivial} and assign bit~1 to it
if and only if the corresponding syndrome bit in round $\tau$ is different from that in round $\tau -1$.
(All syndrome bits in round 0, by definition, are zero.)
That is, a vertex of $V_\tau$ records the \emph{change} in the syndrome bit.
We assign bit~1 to the vertex $b$ 
if and only if the number of nontrivial vertices in $V \backslash\{b\}$ is odd.

Denote by $\mathcal F$ the union of all possible faults (see \secref{sec:noisemodel}) 
that afflict individual elementary operations 
in our optimized  space-time syndrome-extraction circuit (see \figref{fig:schedule}). 
Our circuit has been designed in such a way that any fault in $\mathcal F$ 
either causes only trivial syndrome bits, 
or flips the triviality of exactly two vertices, between which there is an edge in~$E$.
(There are two decoding graphs, one for $X$ syndromes---which is $G$---and the other for $Z$ syndromes, 
and a single fault may flip more than two vertices in total; 
but in each decoding graph the number of flipped vertices is always either zero or two.)
In particular, our circuit induces $\mathcal F$ along with a surjection from $\mathcal F$ to $E$, 
or with a little abuse of notation, a $\mathbb Z_2$-linear map 
$\varphi: \mathbb Z_2^{\mathcal F} \rightarrow \mathbb Z_2^E$.

The Union-Find decoder is fully specified by the decoding graph $G$, 
which is itself determined by the distance~$d$ and the syndrome-extraction circuit.
In a trial of the simulation, the fault configuration can be represented by a subset $F\subseteq \mathcal F$.
The input to the Union-Find decoder is thus the 0-boundary of the 1-chain $\varphi(F)$.
That is, the input is the subset of nontrivial vertices in $V$.   
Then, the decoder will find a subset $C\subseteq E$, whose 0-boundary coincides with the input, in time almost-linear with $|V|$. 
One further projects $C$ into a 1-chain on the two-dimensional spatial plane, 
and each link of this 1-chain corresponds to a (weight-1 or -2) Pauli operator supported on the data qubits. 
The product of all these Pauli operators constitute the final Pauli correction 
(only used by the classical control device by Pauli frame tracking).
The decoding succeeds if and only if $\varphi(F)+C$ is homologically trivial, 
{\em i.e.}, has even overlap with any side of the boundary.
The shortest homologically nontrivial loop in $G$ has length $d$, 
and it follows from~\cite{delfosse2017peeling, delfosse2017unionfind} 
that $\varphi(F)+C$ is trivial as long as $2\,|\varphi(F)|<d$.
Hence the decoding is guaranteed to succeed as long as the number of faults 
$|F|\le\left\lfloor\tfrac{d-1}{2}\right\rfloor$.

\section{Inclusive Error Model}
\label{app:inclusivemodel}

We continue explaining our numerical simulation using the notations introduced in \appref{sec:unionfind}.

Conventional Monte Carlo simulation for error correction  involves \emph{fault sampling}. 
That is, for each elementary circuit operation, one fault is randomly chosen out of a finite set. 
The eventual chosen faults constitute the fault configuration~$F$. 
The overall time taken by each trial is $O(d^3)$. 
Here, due to the graphical nature of the surface code, we instead adopt \emph{edge sampling}. 
Specifically, in each trial we sample the 1-chain $\varphi(F)$ by sampling edges in $E$. 
We pick each edge $e\in E$ independently, whose probability equals the sum of probabilities of those faults that flip $e$. 
The time consumed by edge sampling is still $O(d^3)$, but has favorable constant factor
reduced almost two orders of magnitude since many faults map onto the same edge.

Below we will prove the efficacy of edge sampling for the depolarizing noise model, 
starting with introducing the {\em inclusive} and {\em exclusive} error models.  
The exclusive model refers to the standard stochastic Pauli error model. 
For example, consider a single-qubit gate, which is randomly affected by a Pauli fault $f\in\{I,X,Y,Z\}$ with probability $Q(f)$. 
Note that $\sum_{f \in\{I,X,Y,Z\}}Q(f)=1$, meaning that different faults occur exclusively. 
However, one can also adopt an inclusive model: the gate is first afflicted by $X$ with some probability $P(X)$, then, independently, afflicted by a subsequent $Y$ with probability $P(Y)$, and then $Z$ with probability $P(Z)$. 
The values of $P(X),P(Y),P(Z)$ are within $[0,1]$, and are not constrained otherwise; in particular, they are independent.  
Observe that $P$ and $Q$ satisfy the following relations:
\resizebox{.98\hsize}{!}{
\begin{minipage}{\linewidth}
\begin{align}
\begin{split}
Q(I)&=\overline{P(X)} \cdot \overline{P(Y)} \cdot \overline{P(Z)} + P(X) \cdot P(Y)\cdot P(Z) \\
Q(X)&=P(X) \cdot \overline{P(Y)} \cdot \overline{P(Z)}+\overline{P(X)} \cdot P(Y) \cdot P(Z) \\
Q(Y)&=\overline{P(X)} \cdot P(Y) \cdot \overline{P(Z)}+P(X) \cdot \overline{P(Y)} \cdot P(Z) \\
Q(Z)&=\overline{P(X)} \cdot \overline{P(Y)} \cdot P(Z)+P(X) \cdot P(Y) \cdot  \overline{P(Z)} \label{eqn:PQ2}
\end{split}
\end{align}
\end{minipage}
}
where $\overline{P(f)}=1-P(f)$. 

The above definitions of exclusive and inclusive models for a single-qubit gate can be extended to multi-qubit gates or measurements, as long as the possible faults thereof form a group isomorphic to $\mathbb Z_2^n$ for some integer $n\ge2$.  
For example, the models with $n=2,3,5$ can respectively characterize the cases of  single-qubit identity gate, single-qubit measurement and joint measurement; see \secref{sec:noisemodel}.  

Furthermore, there is a general relation between exclusive model $Q$ and inclusive model $P$, which is analogous to \eqnref{eqn:PQ2}. 
For a general gate or measurement, denote the set of its nontrivial faults by $\mathcal E = \mathbb Z_2^n\backslash \{0^n\}$ with some integer~$n\ge2$. 
An inclusive model $P$ is essentially an arbitrary real-valued function $P: \mathcal E \rightarrow [0,1]$.
Then, $P$ induces a probability distribution $Q_P$ over $\mathbb Z_2^n$ such that
\begin{align}
Q_P(f) = \sum_{ \substack{\mathcal S\subseteq \mathcal E: \\ f = \sum_{s\in\mathcal S} s}} \prod_{s\in\mathcal S} P(s)\prod_{s\in\mathcal E\backslash\mathcal S}\overline{P(s)} \quad \forall f\in\mathbb Z_2^n.  \label{eqn:PQn}
\end{align}
That is, $Q_P$ is the exclusive model induced by $P$. 

A natural question then is that, given a general exclusive model $Q$, 
whether there exists an equivalent inclusive model $P$, {\em i.e.}, such as $P$ induces $Q$.
We claim that for any~$n \ge 2$, there exists an exclusive model not induced by any inclusive model.
It suffices to consider $n = 2$, because any distribution with $n=2$ 
is the marginal distribution of some distribution with $n \ge 2$.
Solving \eqnref{eqn:PQ2} for $P$, we have
\begin{align*}
\resizebox{1\hsize}{!}{$
P( f_1 )  = \frac12 \pm \sqrt{ \frac{ \big( \frac12- Q( f_1 ) - Q( f_2 ) \big)  \big (\frac12 - Q( f_1 ) - Q( f_3 ) \big ) }{ 1- 2 Q( f_2 ) -2 Q( f_3 ) } } 
$}
\end{align*}
where $\{f_1, f_2 , f_3 \} = \{ X,Y,Z \}$.
For some choice of $Q$, this solution may not even be real valued. 

However, in the special case where $Q(f)$ is small and uniform over all nonzero $f$,
there always exists a corresponding $P$ that induces $Q$. 
\begin{claim}\label{c:equivalent}
Given $n\ge2$, let $Q$ be an exclusive model such that 
for all $f\neq 0^n$, we have $Q(f) = q \le 2^{-n}$ for some constant $q$.
Consider the inclusive model $P$ defined by 
$P(f)\equiv\frac12\pm\frac12\left(1-2^nq\right)^{2^{1-n}}$
for all $f \neq 0$.
Then, the inclusive model $P$ induces $Q$.
\end{claim}

Note that when $q = o(1)$, $P(f)$ can be chosen to match~$q$ to the first order.  
 
\begin{proof}
We first solve for $p$ of \eqnref{eqn:PQn} for $f = 0$:
\begin{align}\label{eqn:Q0}
1-(2^n-1)q=\sum_{ \substack{ \mathcal S\subseteq\mathcal E\\0=\sum_{s\in\mathcal S}s}} 
p^{|\mathcal S|}(1-p)^{2^n-1-|\mathcal S|}.
\end{align}
The right-hand-side of \eqnref{eqn:Q0} can be simplified using the following lemma. 

\begin{lemma}[MacWilliams identity~\cite{macwilliams1963weight}]
For any binary linear code on $N$ bits, define
\[
W_C(x,y)=\sum_{u\in C}x^{N-\mathrm{wt}(u)}y^{\mathrm{wt}(u)}
\]
where $\mathrm{wt}$ denotes the Hamming weight.
Then 
\[
W_C(x,y)=\frac{1}{\left|C^{\perp}\right|}W_{C^\perp}(x+y,x-y)
\]
where $C^\perp$ is the dual code of $C$. 
\end{lemma}

It is easy to see that  the right-hand-side of \eqnref{eqn:Q0} 
equals $W_C(1-p,p)$ where $C$ is the Hamming code $[2^n-1,2^n-1-n,3]$.  
Using the fact that the dual of Hamming code has uniform Hamming weight, one easily obtains
\[
1-(2^n-1)q=\frac{1}{2^n}+\left(1-\frac{1}{2^n}\right)\cdot\left(1-2p\right)^{2^{n-1}}
\] 
and hence $p=\frac12\pm\frac12\left(1-2^nq\right)^{2^{1-n}}$. 

To see that \eqnref{eqn:PQn} holds for every $f \neq 0^n$ with $P \equiv \frac12\pm\frac12\left(1-2^nq\right)^{2^{1-n}} $, 
we are going to prove that for any $ f,f'\in\mathcal E = \mathbb Z_2^n \setminus \{0^n\} $ and $1\le k\le 2^n-1$, 
we have $\left|\mathsf A_f^k \right| = \left|\mathsf A_{f'}^k \right|$ where for any $g \in \mathbb Z_2^n$
\begin{align*}
\mathsf A_g^k\; = \; \left\{ \; \mathcal S \; \left| \; \mathcal S\subseteq \mathcal E ,
|\mathcal S|=k \, ,\, \sum_{s\in\mathcal S}s=g \right.\; \right\}.
\end{align*}
This will prove \eqnref{eqn:PQn} 
since the right-hand side of \eqnref{eqn:PQn} is $\sum_k \left| \mathsf A_f^k \right| P^k (1-P)^{2^n-1-k}$.

We find a bijection between $\mathsf A_f^k$ and $\mathsf A_{f'}^k$.
Let $\Delta = f + f' \in \mathbb Z_2^n$ be nonzero.
Since $\mathbb Z_2^n$ is an abelian group, it is partitioned into cosets of a subgroup $\{0, \Delta\}$.
In other words, $\mathcal E = \mathbb Z_2^n \setminus\{ 0 \}$ is partitioned as
\begin{align*}
\mathcal E=\left\{\Delta \right\}\sqcup
\bigsqcup_{ g }\left\{ g, g + \Delta \right\}.
\end{align*}
Consider any $\mathcal S\in\mathsf A_f^k$.
There must exist $g \in \mathcal S$ such that $ g + \Delta \notin \mathcal S$;
otherwise, $f = \sum_{v \in \mathcal S} v$ would be either zero or $\Delta$, 
which is impossible since $f \neq 0$ and $f' \neq 0$.
Fix any total ordering on $\mathcal E$, 
and choose for each $\mathcal S$ the least element $g_{\mathcal S} \in \mathcal S$ such that $g_{\mathcal S} + \Delta \notin \mathcal S$.
Substituting $g$ of $\mathcal S$ with $g+\Delta$,
we have a new subset $\mathcal S' = (\mathcal S \setminus \{ g_{\mathcal S} \}) \cup \{ g_{\mathcal S} + \Delta \}$,
which has exactly $k$ elements.
We thus have a map $\mathcal S \mapsto \mathcal S'$ from $\mathsf A_f^k$ into $\mathsf A_{f'}^k$.
This map is clearly one-to-one.
The mapping from $\mathsf A_{f'}^k$ to $\mathsf A_f^k$ is defined similarly.
\end{proof}

For a syndrome-extraction circuit equipped with an exclusive model $Q$, the edges in $E$ of the corresponding decoding graph are generally not independent. 
Indeed, different faults at a same operation, which are mutually exclusive, may flip different edges.  
In this case, sampling edges independently is not faithful. 

However, if $Q$ admits an equivalent inclusive model $P$, then the events of individual edges being flipped are mutually independent under $P$. 
Specifically, an edge is flipped if and only if an odd number of different faults corresponding to that edge have occurred. 

Recall that $\mathcal F$ is the set of all nontrivial circuit faults in our scheme. 
Our starting noise model in \secref{sec:noisemodel} is an exclusive model $Q$ on $\mathcal F$ where nontrivial faults from a given operation happen uniformly at random.
By \claimref{c:equivalent}, we convert this model to the corresponding inclusive model $P$.
For each $e\in E$, the probability of its being flipped is given by
\begin{align}
\begin{split}
W'(e) &= \sum_{ \substack{F \subseteq \mathcal F_e \\ |F| \textrm{ is odd}} } \prod_{f\in F} P(f) \prod_{f\in \mathcal F_e\setminus F} (1-P(f)) \\
& = \frac12 \left( 1-\prod_{f \in \mathcal F_e}  \left(1-2 P(f) \right) \right) \\
\mathcal F_e &= \left\{ f \, \big | \, f \textrm{ flips }e \right\} \enspace . \nonumber
\end{split}
\end{align}
However, for ease of simulation, we instead calculate the \emph{edge weight}:
\begin{equation}\label{eqn:edgeweights}
W(e) = \sum_{f\in\mathcal F_e} P(f) \enspace .
\end{equation}
Provided that the error rates $P$ are relatively small, independent edge sampling by $W$ is a linear approximation of the conventional fault sampling, which is correct to the second order.

\section{Importance sampling}
\label{sec:importance}

As the code distance increases and the physical error rate $p$ decreases, the event of logical failure becomes so rare that the Monte Carlo simulation is no longer feasible. 
In this section, we explain how to use the importance sampling method to reliably estimate the logical error rate $p_{\rm L}$, {\em i.e.}, the numerical data as in \figref{fig:w_importance} and \figref{fig:d_importance}. 
We will use the notations introduced in \appref{sec:unionfind} and \appref{app:inclusivemodel}.

For simplicity consider only odd distance \mbox{$d=2t+1$}.
Due to the argument about the decoding graph $G=(V,E)$ in \appref{sec:unionfind}, our scheme succeeds whenever there are at most $t$ edges flipped.
Therefore we have
\begin{align*}
\begin{split}
p_{\rm{L}}&=\sum_{w=t+1}^{|E|} A_w \cdot B_w \\ 
A_w&=\Pr \left [ \, w\textrm{ edges flipped} \, \right] \\ 
B_w &=\Pr \left[ \, \textrm{logical failure } \big| \, w \textrm{ edges flipped} \, \right] \enspace. 
\end{split}
\end{align*}
By the argument in \appref{app:inclusivemodel}, given any 1-chain $C\subseteq E$, we calculate from \eqnref{eqn:edgeweights}
\begin{equation}\label{eqn:chainprob}
\Pr\left[ \, C \textrm{ flipped} \, \right] = \prod_{e\in C} W(e) \cdot \prod_{e\in E\setminus C}(1-W(e)) \enspace .
\end{equation}

Our importance sampling method for estimating $p_{\rm{L}}$ goes as follows.
\begin{itemize}[leftmargin=*]
\item Given $t$ and $W$, estimate $w':=\arg\max_w A_w$ by approximating $A_w$ as obeying binomial distribution. \\
Set $N=10^6$ and $I=\{w'',w''+1,\ldots,w''+19\}$, where $w''=\max\{t+1, w'-10\}$.
\item For each $w\in I$ do
\begin{itemize}[leftmargin=*]
\item For $i=1,2,\ldots,N$, sample from $E$ a subset $S_i$ of $w$ edges uniformly at random.
\item $\widehat{A_w}\leftarrow \binom{|E|}{w}\cdot\sum_{i=1}^N \Pr[\, S_i \textrm{ flipped} \,]\,\big/N$, via \eqnref{eqn:chainprob}.
\item For $i=1,2,\ldots,N$, sample from $E$ a subset $S_i$ of $w$ edges by weights $W/(1-W)$, and then decode. 
\item $\widehat{B_w}\leftarrow \sum_{i=1}^N \mathbb I \left[ \, \textrm{logical failure on }S_i \, \right] \big/ N$, where $\mathbb I$ is the indicator function.
\end{itemize}
\item $\widehat{p_{\rm{L}}} \leftarrow \sum_{w\in I}\widehat{A_w}\cdot\widehat{B_w}$
\end{itemize}

It is easy to verify that $\widehat {A_w}$ and $\widehat {B_w}$ converge to $A_w$ and $B_w$ respectively when $N$ goes to infinity. 
In addition, $\widehat {p_{\rm{L}}}$ should be reasonably faithful since $I$ includes typical $w$'s with largest probabilities. 

Due to the vanishing $B_w$ with increasing $t$,  we have only managed to perform the above importance sampling procedure for $t$ up to 6. 
It would be interesting to develop more efficient sampling algorithms for rare events, {\em e.g.}, extending methods in~\cite{bravyi2013simulation} to the Union-Find decoder.

\end{document}